\documentclass[sigconf, nonacm]{acmart}

\usepackage{listings}
\usepackage{xcolor}
\usepackage{enumitem}
\usepackage{booktabs}
\usepackage{tabularx}
\usepackage{array}
\usepackage{longtable}

\AtBeginDocument{%
  }

\begin{document}

\title[Deconstructing Open-World Game Mission Design Formula]{Deconstructing Open-World Game Mission Design Formula: A Thematic Analysis Using an Action-Block Framework}

\author{Kaijie Xu}
\affiliation{%
  \institution{McGill University}
  \city{Montréal}
  \state{Québec}
  \country{Canada}}
\email{kaijie.xu2@mail.mcgill.ca}

\author{Yiwei Zhang}
\affiliation{%
  \institution{McGill University}
  \city{Montréal}
  \state{Québec}
  \country{Canada}}
\email{yiwei.zhang3@mail.mcgill.ca}

\author{Brian Yang}
\affiliation{%
  \institution{McGill University}
  \city{Montréal}
  \state{Québec}
  \country{Canada}}
\email{brian.yang2@mail.mcgill.ca}

\author{Clark Verbrugge}
\affiliation{%
  \institution{McGill University}
  \city{Montréal}
  \state{Québec}
  \country{Canada}}
\email{clump@cs.mcgill.ca}

\begin{abstract}
Open-world missions often rely on repeated formulas, yet designers lack systematic ways to examine pacing, variation, and experiential balance across large portfolios. We introduce the Mission Action Quality Vector (MAQV), a six-dimensional framework—covering combat, exploration, narrative, emotion, problem-solving, and uniqueness—paired with an action block grammar representing missions as gameplay sequences. Using about 2200 missions from 20 AAA titles, we apply LLM-assisted parsing to convert community walkthroughs into structured action sequences and score them with MAQV. An interactive dashboard enables designers to reveal underlying mission formulas. In a mixed-methods study with experienced players and designers, we validate the pipeline’s fidelity and the tool’s usability, and use thematic analysis to identify recurring design trade-offs, pacing grammars, and systematic differences by quest type and franchise evolution. Our work offers a reproducible analytical workflow, a data-driven visualization tool, and reflective insights to support more balanced, varied mission design at scale.
\end{abstract}

\keywords{Open-world Games, Quest Design, Game Analytics, Visualization, Thematic Analysis}

\ccsdesc[500]{Applied computing~Computer games}

\maketitle

\section{Introduction}

In open-world AAA games, the quality of mission design is central to player engagement; however, many recent titles are criticized for ``checklist''-style design that leads to repetition and pacing fatigue rather than sustaining meaningful play. Critics point to large maps populated with formulaic tasks and traversal that functions more like a menu than a world, resulting in a hollow sense of place and grind-like loops \cite{wang2024meaningful}. We argue that beyond taste or production constraints, there is a tooling gap: designers lack a shared, analyzable representation of mission structure and the experiential trade-offs it creates.

Prior work in game design and research provides partial answers. Quest formalisms and patterns (e.g., grammars, archetypes) describe how quests can be composed and varied \cite{smith2011situating, machado2017structure}; debates on ludonarrative tension and models of enjoyment speak to why pacing and rhythm matter for aligning mechanics with story and affect \cite{hocking2009ludonarrative, sweetser2005gameflow}. Meanwhile, game analytics has equipped studios with massive telemetry and dashboards, advancing the visualization of emergent player behavior but leaving a ``semantic leap'' between low-level logs and interpretable design structure \cite{drachen2013game, moura2011visualizing, pirolli2011introduction}. Recent sequence-mining and representation-learning approaches begin to encode player action as language-like sequences, yet still target observed play rather than authored mission blueprints \cite{maram2023mining, wang2024player2vec}. We build on these threads by shifting the analytic unit from player traces to the designer's plan: a grammar of \emph{action blocks} and a six-dimensional experiential lens that makes authored mission flow visible and comparable.

We introduce a two-part framework: (1) the Mission Action Quality Vector (MAQV), which consists of six experiential dimensions (Problem-Solving, Combat, Uniqueness, Exploration, Narrative, Emotion) scored over action sequences, and (2) an action block grammar that normalizes walkthrough text into common blocks (traversal, social, puzzle, stealth, combat, etc.). Using community walkthroughs, we built a cross-title corpus (\(\sim\)2200 missions, 20 open-world games), extracted structured sequences with LLM-assisted conversion, and coupled them with coordinated visualizations (radars, quality timelines, block timelines). All walkthrough sources are public, and our user study received institutional ethics approval with informed consent. Our use of LLMs follows HCI work that positions AI as a collaborator for scalable and interpretable qualitative structuring and sense-making, not an oracle \cite{feuston2021putting, hong2022scholastic, gero2024supporting}.

\begin{figure*}[t]
  \centering
  \includegraphics[width=\textwidth]{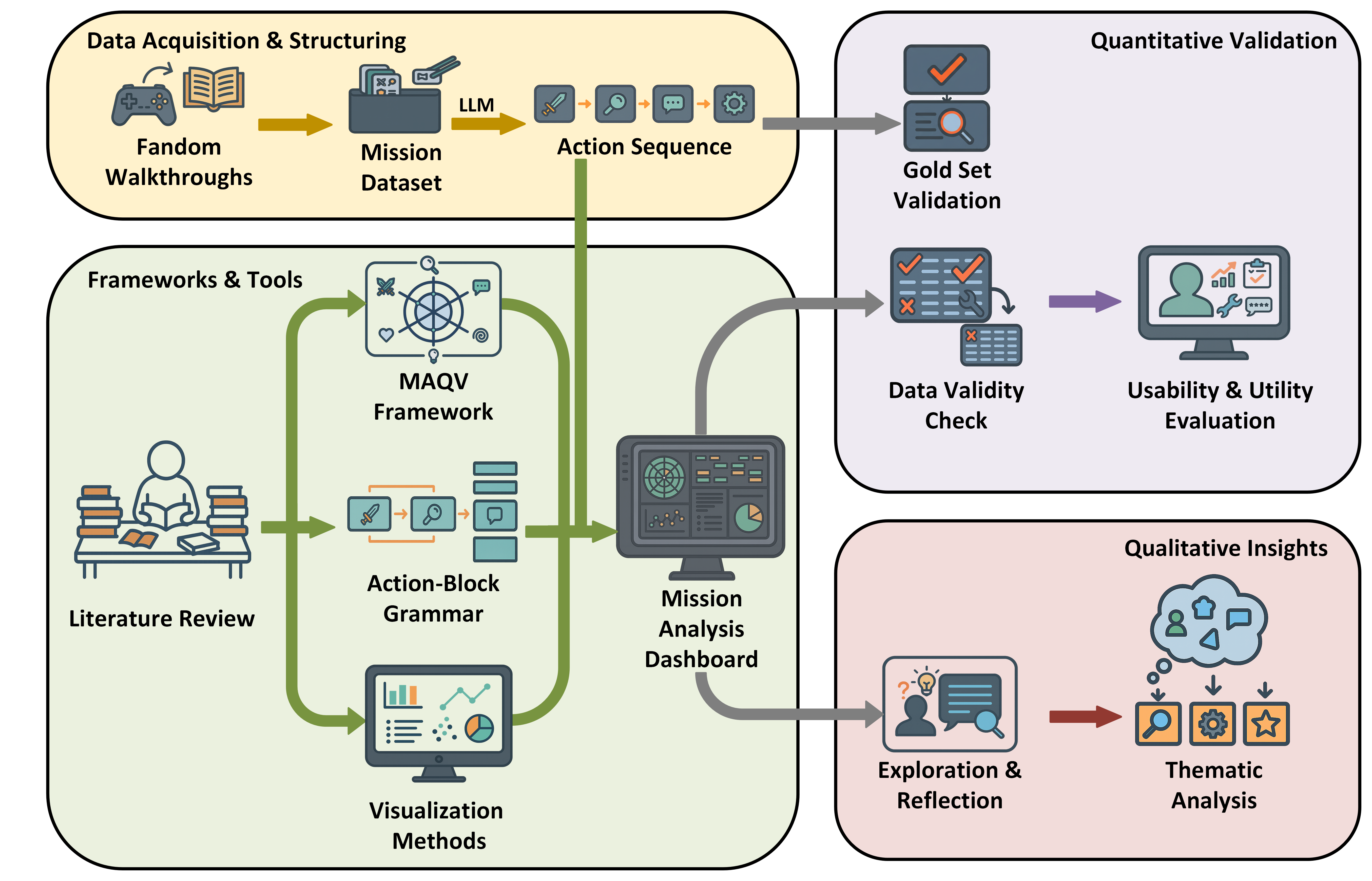}
  \caption{End-to-end workflow of our study. \textbf{Left-top (Data Acquisition \& Structuring):} community Fandom walkthroughs (text only, via MediaWiki API) are consolidated into a mission dataset and converted by an LLM-assisted parser into normalized \emph{action block} sequences. \textbf{Left-bottom (Frameworks \& Tools):} the MAQV six-dimensional lens (combat, exploration, narrative, emotion, problem-solving, uniqueness) and the action block grammar feed a Mission Analysis Dashboard with Browse/Compare views. \textbf{Right-top (Quantitative Validation):} extraction is assessed against a human-labeled gold set, complemented by participant data-validity checks and usability/utility evaluation (SUS, UEQ-S, SEQ). \textbf{Right-bottom (Qualitative Insights):} participants' exploration \& reflection inform Reflexive Thematic Analysis to surface themes, trade-offs, and pacing heuristics.}
  \Description{
    Workflow from data acquisition to validation and qualitative insights in mission design analysis. The diagram has four regions connected by left-to-right flows with feedback links. Top-left (Data Acquisition \& Structuring): community Fandom walkthroughs are consolidated into a mission dataset, then a Large Language Model (LLM) converts text to a normalized action sequence. Center-left (Frameworks \& Tools): the Mission Action Quality Vector (MAQV), an action block grammar, and visualization methods feed a Mission Analysis Dashboard. Top-right (Quantitative Validation): the action sequences are assessed against a human-labeled gold set and a data-validity check, followed by a usability/utility evaluation. Bottom-right (Qualitative Insights): participants' exploration and reflection inform a reflexive thematic analysis.
    }

  \label{fig:workflow}
\end{figure*}

We evaluated the framework in a mixed-methods study with experienced players and professional designers. First, participants verified action lists and extracted sequences, then assessed tool usability using SUS, UEQ-S, and SEQ, following established HCI practice \cite{brooke1996sus,hinderks2017design}.  Next, treating the interface as a reflection probe \cite{wallace2013making}, participants freely explored missions/games and reflected on pacing, archetypes, and cross-title patterns. We analyzed these reflections using Reflexive Thematic Analysis (RTA) \cite{braun2006using}. Separately, we assessed interface usability quantitatively and reported those scores alongside the qualitative themes. The overall workflow is shown in Figure~\ref{fig:workflow}.

Our key contributions are:
\begin{itemize}
    \item a design-facing representation that couples the Mission Action Quality Vector with an action block grammar to externalize mission pacing and experiential balance;
    \item a large cross-title corpus and LLM-assisted pipeline for converting walkthroughs into analyzable sequences; 
    \item a pilot multi-view dashboard that operationalizes MAQV analyses for sense-making over mission structure;
    \item a quantitative evaluation indicating the feasibility of the framework, extraction pipeline, and system;  and
    \item a qualitative study using Reflexive Thematic Analysis that derives actionable pacing heuristics and cross-game design insights.
\end{itemize}
\section{Related Work}

\subsection{Game Analytics and Player Behavior Visualization}

Game Analytics is the process of discovering and communicating patterns in data to improve the player experience and solve design problems \cite{el2016game}. The field traditionally relies on telemetry, a stream of event data transmitted from game clients, to understand player behavior \cite{el2016game, FBK2025}. However, the sheer scale of telemetry data creates a `sense-making gap': prior surveys and systems report abundant signals but limited designer-facing, task-appropriate tooling to map raw metrics to concrete design hypotheses and pacing insights \cite{moura2011visualizing, wallner2013visualization, su2021comprehensive, seif2020data}. As a result, visualization becomes a crucial bridge between raw numbers and intelligible patterns \cite{moura2011visualizing}. Unlike prior work centered on player traces, we target the designer's blueprint of missions to support upstream design reasoning.

Commercial dashboards track KPIs and spatial aggregates with standard charts and heatmaps \cite{el2016game, medler2011data, wallner2013visualization}, while research prototypes visualize complex player processes and strategies. Novel approaches include interactive maps that overlay sequences of player actions over game environments \cite{moura2011visualizing}, and specialized symbolic encoding systems that represent high-level player strategies in a compact, comparable format 
\cite{li2019visualizing, maram2023mining}. These tools are powerful for analyzing emergent player behavior. Our work extends this visual-analytic spirit, but shifts the subject of analysis from the player's path to the designer's blueprint. We adapt similar principles to visualize authored mission structures, addressing the distinct challenge of understanding predefined design formula rather than emergent player tactics.

A fundamental challenge in analyzing telemetry is the ``semantic leap'': transforming raw, noisy event logs into meaningful, interpretable sequences of actions \cite{gomez2020exploring}. Researchers have employed advanced techniques to address this, including process mining to identify common puzzle-solving attempts \cite{gomez2020exploring}, domain-based spatial abstraction to reveal high-level strategies in player movement \cite{maram2023mining}, and even NLP models like player2vec that treat action sequences as a form of language \cite{wang2024player2vec}. Our research takes a complementary path. Instead of mining low-level, high-frequency telemetry, we mitigate the semantic leap by starting with human-authored data from mission walkthroughs. This approach leverages data that is already injected with semantic and narrative meaning, allowing our analysis to focus directly on authored design patterns.

Dashboards are evolving from passive displays to active sense-making tools \cite{munzner2025visualization,sarikaya2018we,heer2012interactive}, yet designers still need customized systems adapted for creative workflows \cite{rocha2021gameplay}. By reframing the dashboard as an interactive environment for inquiry and discovery \cite{sarikaya2018we, sedlmair2012design}, we position our framework as such a tool: a specialized, reflective probe built not for general business intelligence, but for the focused, creative task of deconstructing and innovating in mission design.

\subsection{Quest and Mission Design Literature}

A deep understanding of quest design requires examining the formalisms that structure them and the experiential critiques that challenge them. Quests are the primary mechanism for narrative progression in many genres, guiding players through a story via a series of goals and rewards \cite{machado2017structure}. Researchers have developed several formal models to deconstruct these structures. These include hierarchical planning, where quests are top-level plans decomposed into causally linked events, and more detailed grammar-based models. Pioneering work in this area identified that quests can be generated from a grammar rooted in nine core NPC motivations (e.g., Knowledge, Protection), which expand into strategies (e.g., ``Kill Pests'') composed of atomic player actions (goto, kill, gather) \cite{machado2017structure}. Complementing these generative models are design patterns, such as ``Arrowhead Questing,'' a structure that narrows from broad objectives to specific, high-stakes encounters to teach mechanics and build tension \cite{smith2011situating}. Recent work shows that thematization can shape verb sets and bespoke modes in open-world action-adventure games \cite{junnila2025thematization}. These game-specific formalisms are modern, interactive expressions of timeless literary archetypes like ``The Quest,'' sharing core elements of journey, challenge, and transformation \cite{Hillerich2022QuestPlot, howard2022quests}. Our work builds on this tradition of formal analysis by introducing a taxonomy of ``action blocks'' that deconstructs quests at a more granular, player-action level, to extract the anticipated events players will actually experience through the game.

Underlying quest design is a foundational tension between gameplay and story, crystallized in the ``ludology versus narratology'' debate \cite{mcmanus2006narratology}. Ludology prioritizes the formal systems of rules and mechanics, while narratology focuses on the game as a storytelling medium \cite{veloya2023analysis}. This is not merely an academic argument but a practical design challenge in creating cohesive experiences where play and plot reinforce each other \cite{lindley2005story}. A key technique for resolving this ``ludonarrative'' tension is pacing: the deliberate manipulation of speed and rhythm to control the player's emotional journey \cite{reagan2016emotional}. Effective pacing relies on the cycle of tension and release \cite{cutting2016evolution, sweetser2005gameflow}, which operates on multiple scales, from the moment-to-moment action to the overall arc of the game \cite{shepard2014interactive}. Our Mission Action Quality Vector (MAQV) provides a novel, quantifiable method for visualizing this dynamic, where shifts between dimensions like Combat, Exploration, and Narrative directly represent the designed pacing and tension flow within a quest.

Despite established formalisms, a significant contemporary critique argues that many modern open-world games fail at the experiential level, defaulting to the ``open-world checklist'' problem \cite{wang2024meaningful}. This critique poses that in pursuing scale, games create vast worlds that are fundamentally uninteresting to traverse, reducing activities to icons on a map to be ticked off a list \cite{hughes2023understanding}. The journey becomes ``dead space'' between objectives, bypassed with fast-travel or boring traversal experience, rendering the world a ``massive mission-select screen'' \cite{wang2024meaningful, fernandez2024introduction}. This emptiness is often filled with repetitive, low-value ``filler'' content, like clearing out identical enemy camps, which provides diminishing returns and contributes to a sense of grind. Major franchises, including later Assassin's Creed and Bethesda's RPGs, are frequently cited as exemplars of this design issue \cite{alexander2017deriving, gotz2021evolution}. This critique highlights an important information problem for designers. Our visual analytics framework directly addresses this by making the holistic structure of the player's designed journey visible, enabling designers to move beyond tracking discrete objectives and to analyze whether quest flows create meaningful, engaging paths through their worlds.

\subsection{Methodological Background: Evaluating Creativity Support Tools and Thematic Analysis}

Evaluating Creativity Support Tools (CSTs) is a long-standing HCI challenge because creative work is ambiguous and subjective; conventional metrics (time, errors) are often inadequate \cite{remy2020evaluating,carroll2013quantifying}. The community therefore relies on a methodological ``toolbox'' rather than a single standard \cite{remy2020evaluating}. In this study we evaluate a designer-facing visual analytics system for mission-design sense-making using both quantitative instruments and qualitative inquiry.

Standardized questionnaires offer comparable signals about perceived usability and experience. We use the System Usability Scale (SUS) \cite{lewis2018system} and the short form User Experience Questionnaire (UEQ-S, 8 items; two scales: Pragmatic and Hedonic) \cite{schrepp2017construction,boothe2024generalized,schrepp2014applying}. We did not administer the Creativity Support Index (CSI) \cite{cherry2014quantifying,shneiderman2002creativity} because our evaluation targets interaction-level analysis tasks rather than creative artifact production, and CSI's constructs overlap with UEQ-S. These instruments provide useful but shallow coverage \cite{carroll2013quantifying}, so we benchmark with SUS/UEQ-S and complement them with qualitative methods aligned with expert reflective practice.

To access situated interpretation, we draw on cultural probes and, in particular, the Reflection Probe \cite{wallace2013making,boehner2007hci,loi2007reflective}. We treat the system as a reflection probe and ask whether it helps experts build meaningful representations and make previously intractable analysis possible \cite{pirolli2011introduction,gero2024supporting}. Concretely, our interface externalizes mission structure to support design inquiry, and we study how users reason with these externalizations.

For open-ended reflections, we adopt Reflexive Thematic Analysis (RTA) \cite{terry2017thematic,naeem2023step,bowman2023using,cooper2022systematic,wood2025thematic,devine2021reflexive}. RTA is flexible yet explicit about reflexivity and epistemology: themes are constructed rather than ``discovered,'' and positivist framings (e.g., IRR as a validity claim) are philosophically incompatible with this stance \cite{hallgren2012computing,naeem2023step}. Consistent with this stance, agreement/accuracy statistics (e.g., IRR; precision/recall/F1) are reported only for the quantitative pipeline and are not used to validate qualitative themes. Given reports of inconsistent TA reporting in HCI and game studies \cite{eum2023thematic,rahman2017gamers,vakeva2025don,bowman2023using}, we state our RTA position for coherence and transparency. Our goal is not to derive new explanatory theory but to interpret how players and designers make sense of ``mission formula'' with the tool. RTA suits this aim better than Grounded Theory \cite{glaser1998grounded}. Grounded Theory seeks to build theory through theoretical sampling and saturation with minimal prior structure, but in our study we instead focus on interpreting participants' reflections using existing ideas and their interactions with our dashboard; we do not aim to develop a new theory. Unlike Interpretative Phenomenological Analysis (IPA) \cite{eatough2017interpretative}, RTA supports a somewhat broader expert sample while preserving depth.

Overall, we take a mixed-methods stance: SUS/UEQ-S establish a baseline \cite{lewis2018system,schrepp2017construction,boothe2024generalized,schrepp2014applying}; a reflection-probe evaluates sense-making \cite{wallace2013making,boehner2007hci,loi2007reflective,pirolli2011introduction,gero2024supporting}; and RTA analyzes expert reflections \cite{terry2017thematic,naeem2023step,bowman2023using}. We do so because neither fully ``objective'' nor fully ``subjective'' approaches suffice: widely used objective measures correlate only modestly with perceived quality \cite{hornbaek2007meta}, whereas reflexive analysis cautions against treating accounts as general laws \cite{braun2006using,braun2019thematic}. Accordingly, we triangulate numeric and experiential evidence \cite{denzin2017research}. Procedural specifics of our thematic analysis appear in Section~\ref{sec:thematic_analysis}; here we articulate the methodological foundations motivating our mixed-methods design.
\section{Study 1: Mission Action Quality Vector}
\label{sec:s1}

To reveal the latent ``mission design formula'' that shapes open-world quests, we employ a mixed-methods, data-driven strategy that couples large-scale computational analysis with designer-centered qualitative inquiry. Our study progresses in seven phases. We begin by building a cross-title mission corpus and defining the MAQV, and we conclude with a reflexive thematic analysis. This phased approach allows us to triangulate the quantitative signals with expert reflections, ensuring both breadth and depth of insight \cite{denzin2017research}. Figure~\ref{fig:workflow} overviews our end-to-end pipeline: from dataset construction and LLM-assisted action extraction to visual analytics and mixed-methods evaluation.

We organize the overall workflow into seven phases. Phases 1-5 (MAQV design, dataset construction, action block extraction, gold-set validation, visualization) are detailed in the subsequent subsections of Study~1 (Section~\ref{sec:s1}), while Phases 6-7 (user studies and thematic analysis) are reported in Study~2 (Section~\ref{sec:s2}).

\begin{enumerate}
  \item \textbf{Design of the Mission Action Quality Vector:} We defined a six‐dimensional scoring system to capture the key qualities of every mission action.
  \item \textbf{Dataset construction:} We extracted mission descriptions from community wikis, programmatically parsed and filtered them, and converted each into an ordered sequence of actions grounded in the MAQV taxonomy.
  \item \textbf{LLM‐based Action Block Extraction:} Using LLMs and a predefined action vocabulary, we mapped each mission description into an action sequence, followed by automated validation checks and manual spot reviews.
  \item \textbf{Validation against a human-labeled gold set:} We drew a stratified random sample of 80 missions, obtained independent double annotations and adjudicated consensus sequences, and evaluated LLM outputs via sequence alignment and step-level metrics. 
  \item \textbf{Visual analytics dashboard development:} We implemented an interactive Flask‐based interface with both Browse and Compare modes, integrating diverse visualization methods.
  \item \textbf{User studies with game designers:} We conducted hands‐on sessions in which domain experts used the tool to validate its usability and to generate initial design reflections and hypotheses.
  \item \textbf{Reflexive thematic analysis:} Guided by Braun and Clarke's six‐phase method \cite{braun2006using}, our research team coded the qualitative data from the user studies and the MAQV‐augmented visualizations, iteratively developing and refining themes around mission design insights.
\end{enumerate}

\subsection{Methodology}

In Study~1, we focus on constructing the MAQV representation and the walkthrough-to-sequence pipeline, which together yield MAQV-scored action sequences at scale. We then operationalize these outputs in a dashboard, while quantitative/qualitative evaluations of fidelity and usability are reported in Study~2.

\subsubsection{\textbf{MAQV Framework}}
\label{sec:MAQV}

\begin{table*}[t]
\centering
\small
\setlength{\tabcolsep}{4pt}
\caption{Content validity for MAQV: each dimension is anchored in established game-design constructs and tied to our observable cues derived from action blocks.}
\label{tab:maqv_content_validity}
\begin{tabularx}{\linewidth}{l X X}
\toprule
\textbf{Dimension} &
\textbf{Theoretical anchors (refs)} &
\textbf{Operational cues (from action sequences)} \\
\midrule
Uniqueness (U) &
Novelty as a driver of engagement; need for novelty in SDT/uncertainty frameworks
\cite{tsay2020overcoming,peng2021detecting,to2016integrating,deterding2022mastering,kosa2024exploration}. &
First-appearance flags of mechanics/actions; density of one-off set pieces; rate of introducing new blocks. \\
\addlinespace[2pt]
Combat (C) &
Adversarial challenge and risk in play; combat as an agency-shaping core
\cite{ho2022game,Mossner2022,osborn2023combat,hutchinson2007performing}. &
Share of combat/stealth/escort blocks; intensity markers (boss tags, wave counts, sustained encounters). \\
\addlinespace[2pt]
Narrative (N) &
Quest-mediated story progression; measuring story factors; narrative alignment
\cite{naul2020story,howard2022quests,qin2009measuring,moser2015narrative,hocking2009ludonarrative}. &
Proportion of dialogue/cutscene/exposition blocks; presence of plot turns or quest-stage gates. \\
\addlinespace[2pt]
Exploration (E) &
Discovery motives (Bartle) and MDA ``Discovery'' aesthetic
\cite{bartle1996hearts,hunicke2004mda}. &
Share of traversal/scouting/collection blocks; POI/landmark discovery events. \\
\addlinespace[2pt]
Problem-Solving (P) &
Games as problem spaces; puzzle/strategy and cognitive load/flow
\cite{juul2003game,jorgensen2003problem,liu2011effect,fullerton2024game}. &
Share of puzzle/investigation/manipulation blocks; multi-step precondition chains. \\
\addlinespace[2pt]
Emotion (A) &
MDA ``Aesthetics'', Four Keys to Fun, and affect in play
\cite{hunicke2004mda,games2004four,Hemenover2018,karpouzis2016emotion,melo2014}. &
Alternation density of high-tension vs.\ relief segments; timed/escape/culmination flags. \\
\bottomrule
\end{tabularx}
\vspace{-2mm}
\end{table*}

To operationalize our analysis of the mission formula, we introduce MAQV, a six-dimensional framework for quantifying the experiential qualities of any given action within a mission. Each action, extracted from mission sequences, is scored from 0 to 1 across six dimensions, creating a vector that represents its qualitative profile. We denote the six MAQV dimensions as Uniqueness (U), Combat (C), Narrative (N), Exploration (E), Problem-Solving (P), and Emotion (A), where A denotes Emotion in the sense of affect/aesthetics. We use the shorthand \textbf{U/C/N/E/P/A} throughout. This vectorization allows us to transform descriptive mission walkthroughs into rich time-series data, enabling computational analysis and visualization of mission pacing, structure, and experiential focus. Each dimension is grounded in established game studies literature to ensure a comprehensive and theoretically sound model of player experience (see Table~\ref{tab:maqv_content_validity}). Raters applied a concise, anchored 0-1 rubric per dimension: 0 denotes absent or generic presence; 0.5 indicates moderate, context-supported presence; 1 reflects salient, defining presence. Each dimension was scored independently based on the intrinsic properties of the action (not narrative payoff), with borderline cases defaulting to mid-range ($\approx$ 0.5) and adjusted during the structured consensus review. The six axes form a literature-grounded minimal set; they are not exhaustive, and the schema supports adding/replacing axes or genre-specific weighting when appropriate.

The \textbf{Uniqueness} dimension captures the novelty and surprise of an action. This is informed by research on the ``novelty effect,'' where initial engagement is high but fades with familiarity \cite{tsay2020overcoming, peng2021detecting}. Recent work within Uncertainty and Self-Determination Theory has even proposed the need for novelty as a distinct psychological factor that uniquely predicts positive player experiences and continued engagement \cite{to2016integrating, deterding2022mastering, kosa2024exploration}. Thus, higher scores on this dimension indicate rare or signature mechanics, whereas lower scores denote common or repeated actions.

\textbf{Combat} measures the intensity and frequency of direct, kinetic conflict. Rooted in game theory's focus on adversarial scenarios and strategic decision-making \cite{ho2022game}, this dimension represents a fundamental challenge in many games  \cite{Mossner2022}. Research highlights that combat mechanics centrally affect player agency and experiential dynamics \cite{osborn2023combat, hutchinson2007performing}, distinguishing physical opposition from other challenge forms.

The \textbf{Narrative} dimension evaluates an action's contribution to advancing the story or developing characters \cite{naul2020story}. Quests are a primary bridge between gameplay and narrative \cite{howard2022quests}, and this dimension measures an action's role in maintaining that connection. Prior studies investigate methods of quantifying narrative factors within games \cite{qin2009measuring, moser2015narrative}. Furthermore, this dimension directly addresses the concept of ludonarrative harmony, where gameplay mechanics reinforce narrative themes, as opposed to ludonarrative dissonance, where these elements conflict \cite{hocking2009ludonarrative}. 

\textbf{Exploration} quantifies an action's role in spatial discovery and environmental interaction \cite{si2017initial}. This aligns directly with the motivations of the ``Explorer'' player type in Bartle's taxonomy, who derives pleasure from discovering new areas and secrets \cite{bartle1996hearts}. It also maps to the ``Discovery'' aesthetic in the Mechanics-Dynamics-Aesthetics (MDA) framework \cite{hunicke2004mda}. Thus, actions promoting expansive or secretive environments score higher, while linear or constrained tasks score lower.

\textbf{Problem-Solving} measures the cognitive challenges, including puzzles, strategic thinking, and complex decision-making. This dimension draws from perspectives viewing games as problem-solving systems where overcoming ``aporias''—or barriers—is integral to player progression \cite{juul2003game, jorgensen2003problem, liu2011effect}. It thus directly assesses cognitive load, critical to puzzle design and the maintenance of optimal player flow \cite{fullerton2024game}.

Finally, the \textbf{Emotion} dimension captures the affective engagement an action evokes in the player. This operationalization directly corresponds to the ``Aesthetics'' component of the MDA framework, which emphasizes emotional responses such as challenge, fantasy, or sensation \cite{hunicke2004mda}. Similarly, Lazzaro's Four Keys to Fun framework categorizes play experiences by their emotional impact \cite{games2004four}, such as ``Hard Fun'' (overcoming adversity) or ``Easy Fun'' (driven by curiosity). While previous research has explored multimodal physiological measurements of emotion in games \cite{karpouzis2016emotion}, our approach relies primarily on psychology-based interpretative frameworks \cite{melo2014, Hemenover2018}.

\begin{figure*}[t]
    \centering
    \includegraphics[width=\linewidth]{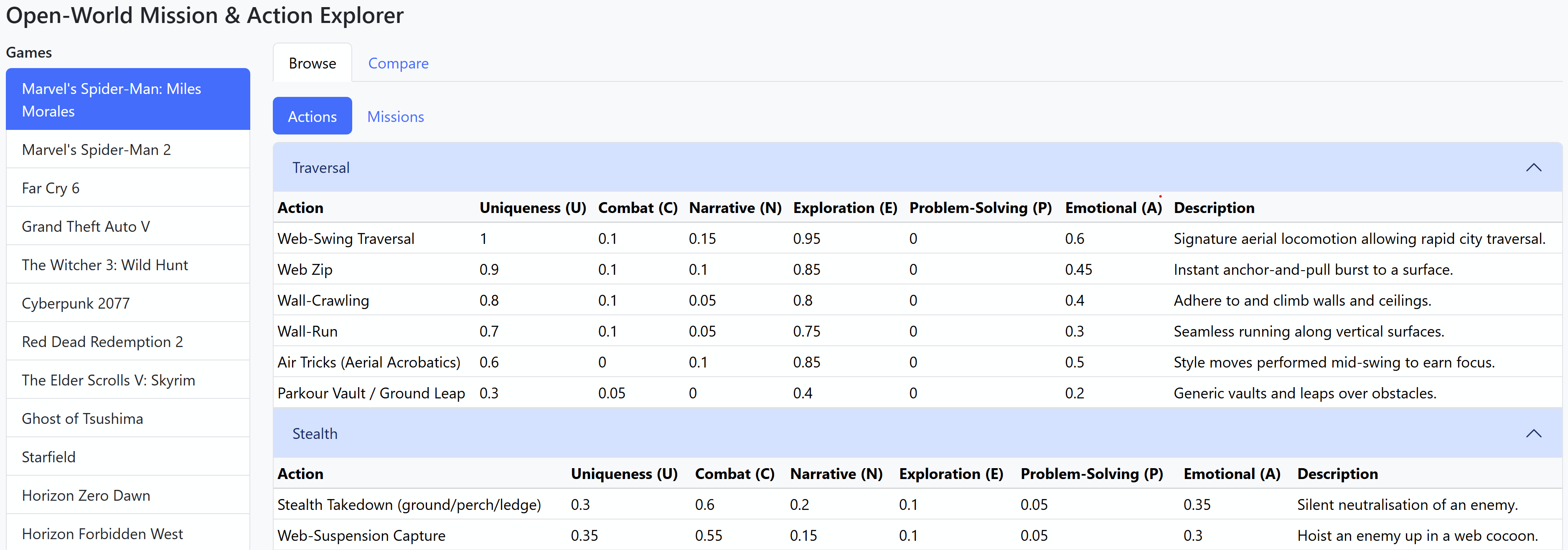}
    \caption{Action view in Browse mode from our implemented tool. Actions are grouped by category; each row shows 0-1 MAQV scores: Uniqueness (U), Combat (C), Narrative (N), Exploration (E), Problem-Solving (P), and Emotion (A), plus a brief description. This figure and all other visualizations in the paper are direct screenshots of the working interface.}

    \Description{Action Explorer—Browse view listing per-game actions with MAQV scores and brief descriptions. The left sidebar lists games with the selected title highlighted; tabs switch between Browse/Compare and Actions/Missions. The main panel shows category sections (e.g., Traversal, Stealth) containing tables of actions. Each row reports 0-1 scores for six Mission Action Quality Vector (MAQV) dimensions—Uniqueness (U), Combat (C), Narrative (N), Exploration (E), Problem-Solving (P), and Emotion (A)—plus a short textual description.}

    \label{fig:action-view}
\end{figure*}

Collectively, these six dimensions offer an integrated yet minimally redundant analytical lens for mission design. They intentionally map onto established theoretical models, including Bartle's player types and MDA aesthetics, ensuring broad coverage of critical experiential facets. For example, both Combat and Problem-Solving pertain to MDA's ``Challenge'' aesthetic but crucially differentiate physical, reflex-based difficulties from cognitive, strategic challenges. Likewise, Narrative measures an action's structural narrative contribution, whereas Emotion assesses immediate affective impacts on the player. This carefully selected dimensionality facilitates refined modeling of mission design tensions, such as between narrative coherence and exploratory freedom. We deliberately omit a Social axis because our analysis targets authored single-player mission flows and the walkthrough corpus encodes solitary play; multiplayer coordination and social dynamics are beyond our present scope. For a concrete illustration using our tool, Figure~\ref{fig:action-view} shows that Web-Swing Traversal in \emph{Spider-Man: Miles Morales} is high on Uniqueness and Exploration (with only minimal Problem-Solving or Combat), whereas Stealth Takedown concentrates on Combat and contributes little to Narrative or Exploration.

Quantification on a continuous 0-1 scale further supports comparative and quantitative analytics. This practice aligns with established methods in game analytics, employing granular telemetry to generate quantifiable metrics, such as completion times or failure rates, to compare engagement across game levels \cite{drachen2013game}. By encoding qualitative gameplay aspects numerically, the MAQV framework allows systematic computational comparison of design patterns across numerous missions and multiple game titles, forming a robust empirical foundation for our analysis.

\subsubsection{\textbf{Dataset}}
\label{sec:dataset}
To enable large-scale, cross-game analysis of mission design, we built a dataset of quest descriptions from 20 major open-world titles. We used Fandom wikis as the primary source: community-curated, publicly accessible repositories that are continuously validated by dedicated player communities. Using online player forums/wikis as research corpora is established in game studies \cite{griffiths2014online}, allowing analysis of experiences, beliefs, and emotions that would be infeasible to collect manually \cite{rautalahti2019video,phillips2021identifying,vakeva2025don}. Before automation, we first assembled a broad candidate pool by surveying public sources (editorial roundups, platform charts, award shortlists, community threads), and then two authors manually vetted each candidate game's public wiki to confirm mission-by-mission coverage sufficient for parsing. Inclusion criteria covered maturity, completeness, and structural consistency so that quest pages were sufficiently detailed and followed a predictable HTML structure suitable for programmatic parsing.

Our final selection comprises 20 critically and commercially successful open-world games across multiple sub-genres (fantasy, science fiction, contemporary) and studios, spanning 2011-2025. The set includes foundational titles that codified genre tropes and recent releases that iterate upon them, enabling a broad analysis of the mission formula. Table~\ref{tab:game_dataset} lists the games, developers, and years.

\begin{table}[t]
\centering
\caption{The 20 open-world games selected for our mission dataset, with developer and year.}
\label{tab:game_dataset}
\begin{tabular}{@{}lll@{}}
\toprule
\textbf{Game Title} & \textbf{Developer} & \textbf{Year} \\
\midrule
The Elder Scrolls V: Skyrim & Bethesda Game Studios & 2011 \\
Dragon's Dogma & Capcom & 2012 \\
Grand Theft Auto V & Rockstar North & 2013 \\
Dying Light & Techland & 2015 \\
Fallout 4 & Bethesda Game Studios & 2015 \\
The Witcher 3: Wild Hunt & CD Projekt Red & 2015 \\
Horizon Zero Dawn & Guerrilla Games & 2017 \\
Kingdom Come: Deliverance & Warhorse Studios & 2018 \\
Red Dead Redemption 2 & Rockstar Games & 2018 \\
Ghost of Tsushima & Sucker Punch Productions & 2020 \\
Spider-Man: Miles Morales & Insomniac Games & 2020 \\
Cyberpunk 2077 & CD Projekt Red & 2020 \\
Far Cry 6 & Ubisoft Toronto & 2021 \\
Horizon Forbidden West & Guerrilla Games & 2022 \\
Hogwarts Legacy & Avalanche Software & 2023 \\
Marvel's Spider-Man 2 & Insomniac Games & 2023 \\
Starfield & Bethesda Game Studios & 2023 \\
Dragon's Dogma 2 & Capcom & 2024 \\
Rise of the Ronin & Team Ninja & 2024 \\
Kingdom Come: Deliverance 2 & Warhorse Studios & 2025 \\
\bottomrule
\end{tabular}
\end{table}

We programmatically retrieved quest descriptions from Fandom via the MediaWiki API and parsed walkthrough-style sections, removing boilerplate. To ensure quality and reproducibility, we retained only descriptions whose cleaned, complete walkthrough description contained at least 35 words and conducted all analyses on immutable page snapshots identified by revision/oldid. This yielded 2435 raw descriptions; after conversion to action sequences (Section~\ref{sec:actionblock}) and quality filtering, the final dataset comprises 2191 valid missions: 607 Main Quests, 1174 Side Quests, and 410 Points of Interest (POIs). Fandom text is licensed under CC BY-SA 3.0; we accessed content via the API in line with the Terms of Use, processed text only (no media), and will release derived annotations (action sequences and per-mission MAQV vectors) with per-page attribution rather than verbatim text. No personal data is involved, and our protocol was reviewed by the IRB. Details of the scraping and parsing pipeline, together with compliance boundaries, are provided in Appendix~\ref{app:data}.

\subsubsection{\textbf{Action Block Extraction}}
\label{sec:actionblock}

We define an action block as an atomic, player-performed verb that is (i) observable in textual walkthroughs, (ii) locally complete and temporally bounded, and (iii) semantically stable enough to compare across missions. To ensure consistent granularity, we organize blocks under a small but general set of categories (e.g., traversal, combat, stealth, puzzle, social interaction, environmental interaction, special ability, etc.). During list construction, we merge near-synonyms (e.g., wall-running vs. wall-crawling if not design-significant), collapse UI-only or meta actions unless they constitute core mechanics, and keep franchise-specific verbs when they meaningfully alter play (e.g., web-swing traversal).

Before LLM extraction, the first author compiled a per-game action block list, the set of unique actions likely to appear in that game's missions, by skimming representative walkthroughs to seed candidates and applying the inclusion/merging rules above. Each action was initially scored on all six MAQV dimensions by a primary annotator with first-hand play experience across the dataset. For each game, one or two secondary annotators who had personally played the title independently reviewed the action list and scores; when no secondary had prior experience, a reviewer completed targeted play before review. Thus, every title had 2-3 annotators with first-hand play. An additional co-author occasionally served as a non-scoring sanity reviewer on titles they had played. We used a consensus protocol similar to a modified Delphi method \cite{diamond2014defining}, which we refer to as a `structured consensus meeting': unanimous ratings were accepted; disagreements were discussed until consensus. This follows expert consensus methods known to improve validity through iterative feedback and convergence \cite{gnatzy2011validating,khodyakov2023rand,schifano2025delphi}. The structured consensus ensured that MAQV assignments reflected shared interpretive understanding and experiential expertise. 

Before the full annotation pass, we ran a pre-annotation inter-rater reliability (IRR) check on two representative titles (The Witcher 3; GTA V) \cite{hallgren2012computing}. To maximize comparability, each MAQV dimension was discretized to a five-point grid {0, 0.25, 0.5, 0.75, 1} for this IRR only. Two authors independently scored 76 actions; quadratic-weighted Cohen's $\kappa$ indicated strong overall agreement ($\kappa$=0.913, 95\% CI [0.895, 0.929]) and moderate-to-excellent agreement per dimension (U 0.737; C 0.966; N 0.868; E 0.953; P 0.890; A 0.843). Exact agreement was 0.767 with 0.228 off-by-one (mean absolute difference 0.059 on the 0-1 scale); see Appendix~\ref{app:irr} for full tables. We then reverted to the continuous 0-1 scale for the main study and finalized scores via the structured consensus procedure above. For clarity, the above IRR serves solely as a procedural reliability check for MAQV action scoring; no coder-agreement statistics were computed for our Reflexive Thematic Analysis.

To convert the unstructured mission descriptions from our dataset into structured, analyzable sequences of actions, we employed a Large Language Model (LLM). Manually coding over 2000 detailed mission texts would be prohibitively time-consuming and prone to inconsistency. Recent work in HCI has demonstrated the utility of AI assistants in qualitative research \cite{feuston2021putting, jiang2021supporting, hong2022scholastic}, capable of performing structured data extraction and aiding in the analysis of large text corpora with significant accuracy \cite{vakeva2025don, kim2025structured, yang2025unified}. We therefore framed this task as a zero-shot, constrained data extraction problem, leveraging the model's ability to parse natural language and map it to a predefined schema. While this approach offers scalability, we did not take the model's output for granted; the validity of the generated action sequences was established through (i) an objective validation study (Section~\ref{sec:validation}) and (ii) convergent participant ratings from the Stage~1 data-validity check (Section~\ref{sec:quant_method}), jointly demonstrating the effectiveness of our extraction pipeline. We report the corresponding fidelity metrics, along with other quantitative evaluation results, in Study~2.

Our process was managed by a programmatic pipeline designed to ensure consistency and robustness. For each of the 2191 viable missions, we dynamically constructed a prompt for the LLM (GPT-4.1, with the fixed snapshot \textit{gpt-4.1-2025-04-14}). This prompt contained four key elements: (1) a system message defining the model's role as a data converter; (2) the complete, game-specific library of predefined actions it was allowed to use; (3) a clear, step-by-step instruction set telling the model to read the description, break it into elementary steps, map each step to exactly one predefined action, and output only a JSON array of action names; and (4) the mission description text itself. To promote deterministic output, we used deterministic decoding (temperature=0, top\_p=1). The pipeline included a robust retry mechanism with a rate limit to handle transient API failures and automatically validated that the model's output was a correctly formatted JSON array of strings. Complete prompt specifications are provided in Appendix~\ref{app:llm_prompt}.

This extraction process successfully generated action sequences for all missions. Our comprehensive action libraries for the 20 games contained a total of 711 manually defined actions, of which 701 were utilized by the LLM in at least one mission sequence, validating the breadth of activities represented. The resulting sequences varied significantly in length, with an average of 12.28 actions per mission. An initial analysis of these sequences reveals clear structural differences between mission types: Main Quests are the longest on average (15.67 actions), followed by Side Quests (11.71 actions), and POIs are the most compact (8.79 actions). This structured dataset of action sequences forms the foundation for the quantitative analysis of mission formula presented in the following sections.

\subsubsection{\textbf{Validation Study: Human-Labeled Gold Set}}
\label{sec:validation}

\begin{figure*}[t]
    \centering
    \includegraphics[width=0.95\linewidth]{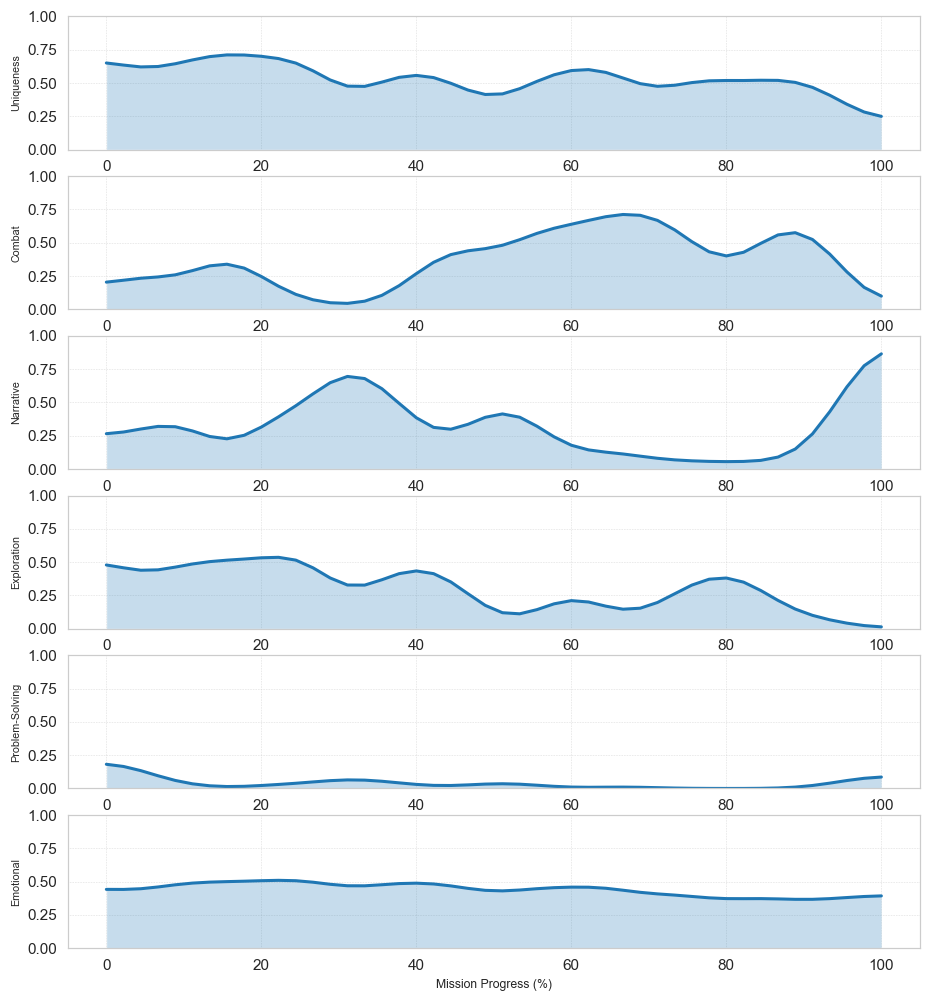}
    \caption{Quality-flow visualization tracking the six MAQV dimensions across normalized mission progress (not wall-clock time). Peaks highlight intense phases; valleys denote lulls; example shown is from \textit{Marvel's Spider-Man 2} (Main Quest: ``A Second Chance'').}
    \Description{Line chart with six stacked panels showing how different mission quality dimensions vary over mission progress from 0 to 100 percent. The first panel, Uniqueness, starts from around 0.7, fluctuates between about 0.4 and 0.6 with multiple small peaks, and finally drops to 0.25. The second panel, Combat, starts low around 0.25, firstly rises to 0.4 and drops to 0.05 before 40 percent, then rises steadily after 40 percent, peaks near 0.7 between 60 and 80 percent, then declines to 0.1. The third panel, Narrative, rises to around 0.7 near 30 percent, dips mid-mission, and rises again sharply to 0.9 at the very end. The fourth panel, Exploration, stays between 0.3 and 0.5 before 50 percent, then drops to around 0.25, with modest rises near 80 percent and gradual drops to 0. The fifth panel, Problem-Solving, remains near zero for most of the mission, with only slight bumps at the start and end. The sixth panel, Emotional, stays relatively flat around 0.5 with minimal variation. Together, the visualization shows that Combat and Narrative intensity increase strongly toward the later stages, while Problem-Solving remains minimal.}
    \label{fig:flow}
\end{figure*}

To objectively assess the reliability of our LLM-based extraction, we conducted a validation study on a stratified random sample of 80 missions drawn from our corpus (20 games; 24 Main, 36 Side, 20 POI). We stratified over game~$\times$~mission-type and sampled with a fixed pseudorandom seed (seed = 42), enforcing at least one mission per type per game where available; if a game lacked a given type, its quota was reallocated proportionally to others. Two annotators (two authors) independently segmented each walkthrough into elementary steps and mapped each step to exactly one item from the closed, per-game action list; a third adjudicator (the first author) produced a consensus gold set and logged disagreements (missing, spurious, mislabel, merge/split, granularity drift).

\begin{figure*}[t]
    \centering
    \includegraphics[width=0.95\linewidth]{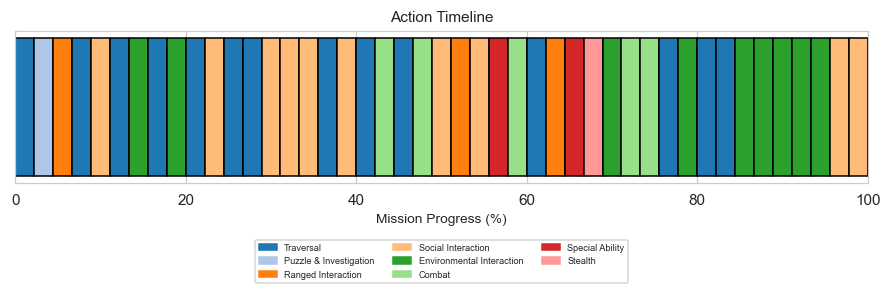}
    \caption{Action-timeline visualization showing the ordered sequence of action categories, enabling rapid assessment of pacing and structural rhythm; example shown is from \textit{Marvel's Spider-Man 2} (Main Quest: ``A Second Chance'').}
    \Description{Bar timeline chart labeled Action Timeline. The horizontal axis shows mission progress from 0 to 100 percent. The bar is divided into many vertical colored segments, each representing a consecutive action category. Categories include: Traversal (blue), Puzzle and Investigation (light blue), Ranged Interaction (dark orange), Social Interaction (peach), Environmental Interaction (green), Combat (dark green), Special Ability (red), and Stealth (pink). Segments are distributed throughout the mission: early sections alternate between Traversal, Social Interaction, and Environmental Interaction; mid-mission sections introduce more Social Interaction,Special Abilities and Combat segments; later sections mix Environmental Interaction, Social Interaction and Traversal; and there are occasional Ranged Interaction and pink Stealth segments in the middle. The visualization conveys pacing by showing how categories alternate over time, illustrating rhythm and structural variation in the mission.}

    \label{fig:action-timeline}
\end{figure*}

LLM-predicted sequences were aligned to gold via Needleman-Wunsch dynamic programming on action labels. We report micro Precision, Recall, and F1, exact sequence match rate, and normalized edit distance, with 95\% bootstrap confidence intervals. Group differences across mission types and across games were tested using non-parametric procedures (Kruskal-Wallis with Holm-corrected post-hoc tests).

\subsection{Visualization Results}

To motivate our choice of a compact set of familiar views for this pilot dashboard, we briefly situate them within prior game analytics visualizations. Prior HCI and games research has used visual analytics to make sense of complex play data, mission structure, and player behavior. Early work includes Medler et al.'s ``Data Cracker'' dashboard for monitoring player metrics via interactive charts and maps \cite{medler2011data}. Hullett and Whitehead analyzed FPS mission design patterns, showing the value of decomposing missions into components and motivating tools that visually map structure \cite{hullett2010design}. Wallner and Kriglstein's survey highlights heatmaps, timelines, and node-link diagrams for revealing behavior patterns and traversal paths \cite{wallner2013visualization}. Building on this, later systems added interactive sequence analysis and comparison; for example, Wallner's Play-Graph visualized gameplay event sequences to expose recurring patterns \cite{wallner2013play}. Dimensionality-reduction (e.g., PCA) has also been applied to player and content telemetry to cluster behaviors or mission types \cite{wallner2013visualization}. More recently, Kepplinger et al. combined timeline, spatial, and physiological views (e.g., synchronized gaze paths and emotion timelines) to support holistic assessment \cite{kepplinger2020see}. Collectively, this work shows that diverse visualizations, from radar/spider charts to flow timelines, help designers analyze and reflect on complex missions in both commercial and research settings \cite{drachen2013game}.

In our methodology, we developed a web-based visual analytics interface that incorporates multiple coordinated visualization techniques to interrogate the ``mission formula'' data from AAA open-world titles. The interface was implemented as a Flask web application (with a Bootstrap front-end) generating charts server-side using Matplotlib and Seaborn libraries. Two main modes of interaction are supported: a \textit{Browse} mode for within-game exploration (per-game and per-mission views), and a \textit{Compare} mode for cross-game visual comparisons. Users can interactively explore patterns, export visualization snapshots, and use the visuals to inform guided design reflection prompts. We adopt a compact set of familiar views to cover three analysis families: emphasis, similarity/contrast, and sequencing/motifs. All views share consistent MAQV encodings and respond to the same subset filters; selecting a title or mission updates panels that summarize that subset. Interactions are intentionally lightweight to keep first-use cognitive load low in this pilot implementation, whose goal is to materialize the analytical workflow rather than introduce novel chart types. 

\begin{figure*}[t]
  \centering
  \includegraphics[width=\linewidth]{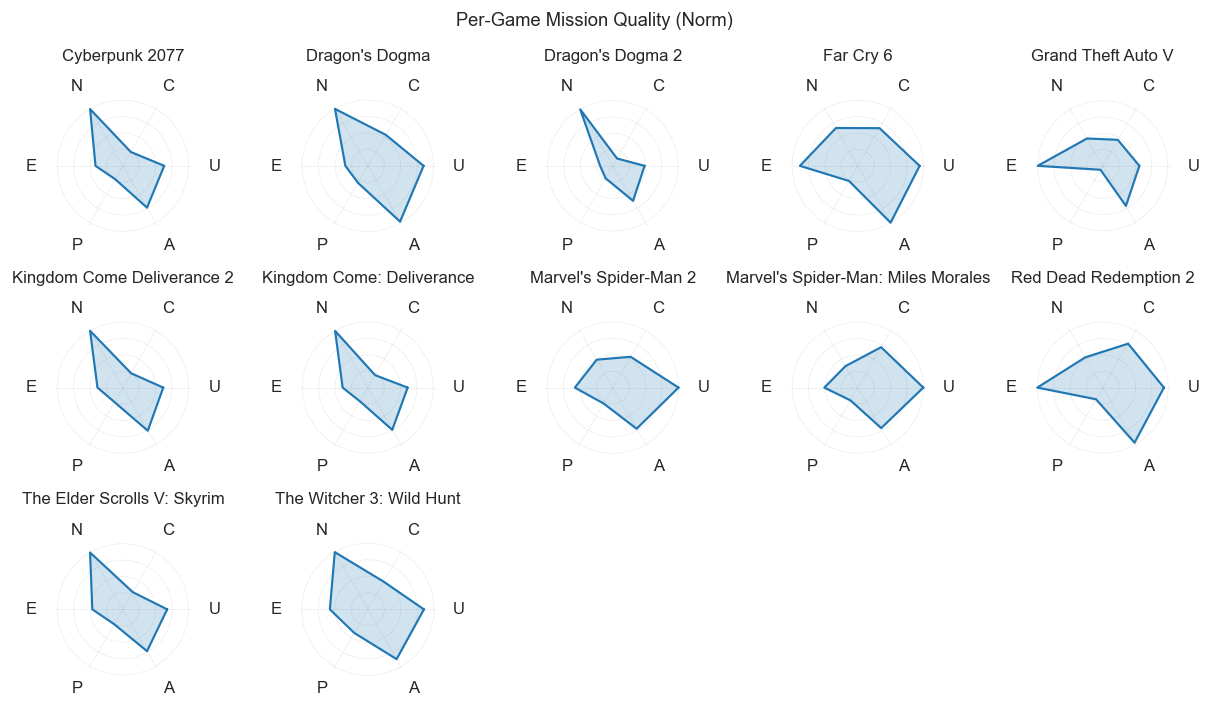}
  \caption{Normalized per-game radar charts highlighting each title's internal balance. U=Uniqueness, C=Combat, N=Narrative, E=Exploration, P=Problem-Solving, A=Emotional.}
  \Description{
    Grid of twelve small radar charts titled ``Per-Game Mission Quality (Norm)''.
    Each chart is normalized within a game and uses six axes: U=Uniqueness, C=Combat, N=Narrative, E=Exploration, P=Problem-Solving, A=Emotional. Polygons show relative emphasis rather than absolute cross-title magnitudes. Cyberpunk 2077: strong N and A; moderate U; low E and C; P near zero. Dragon's Dogma: very strong A and N with high U; modest C; low E and P. Dragon's Dogma 2: high N; moderate U and A; low C and E; P minimal. Far Cry 6: high U and A with notable E; N and C around mid-range; P low. Grand Theft Auto V: dominant E; moderate N and A; moderate U; low C; P minimal. Kingdom Come: Deliverance 2: high N; mid U and A; low E and C; P minimal. Kingdom Come: Deliverance: high N; mid U and A; lower E and C; P minimal. Marvel's Spider-Man 2: highest on U; mid C, A, and E; lower N; P minimal. Marvel's Spider-Man: Miles Morales: similar profile—U highest, mid C/A/E, lower N; P minimal. Red Dead Redemption 2: very strong E and A; high U and C; lower N; P minimal. The Elder Scrolls V: Skyrim: high N and A; mid U and E; low C; P minimal. The Witcher 3: Wild Hunt: high U and A with strong N; mid E; low C; P low-to-mid. Overall, the grid highlights per-title internal balance: several games de-emphasize P and C, while N, A, and U vary most by title.}
  \label{fig:mission-radar-per-game}
\end{figure*}

\textbf{Browse mode} supports within-mission reading at two granularities. \emph{Quality-flow} (Figure~\ref{fig:flow}) shows six MAQV traces aligned to mission progress, revealing build-ups, plateaus, and climaxes; smoothing (Gaussian, $\sigma{=}2$) is for visualization only—statistics use raw values. The companion \emph{action timeline} (Figure~\ref{fig:action-timeline}) encodes the same mission as a color-coded sequence of action categories, surfacing macro-structure (e.g., traversal $\leftrightarrow$ stealth alternation, single-peak boss segments). Shared colors and normalized progress \([0,100]\%\) make it easy to link what happens to how it feels \cite{steinberger2011context, doerr2024visual, van2023survey, guo2018visual}. Additional Browse panels (storyboard, per-mission numerical summaries, and the per-game \emph{Action view} table of actions with MAQV scores as shown in Figure~\ref{fig:action-view}) are described in Appendix~\ref{app:vis}.

\textbf{Compare mode} aggregates across selected titles. \emph{Normalized per-game mission radars} (Figure~\ref{fig:mission-radar-per-game}) emphasize shape (within-title balance) so atypical profiles stand out, while \emph{Top Categories / Game} (Figure~\ref{fig:top-categories}) highlights dominant or under-served mechanics. These answer ``which dimensions does this title lean on?'' and ``which gameplay atoms dominate?'' respectively. More generally, Compare mode structures analysis into four question families: Emphasis (radars; Top Categories), Contrast (combined radar means; PCA similarity map; distance matrix), Structure/Outliers (dendrogram; centroid tables), and Motifs (frequent 3-step blocks), with details in Appendix~\ref{app:vis}. All summaries are computed consistently from the same dataset, ensuring stable values across subsets and exports.
\section{Study 2: User Study}
\label{sec:s2}

Study~2 evaluates the fidelity and practical usefulness of our MAQV-based pipeline and Mission Analysis Dashboard through a three-stage mixed-methods study with players and designers. Section~\ref{sec:study2_method} reports the study methodology, including participants, a quantitative component (Stage~1 data validity check; Stage~3 usability \& utility instruments), and a qualitative component (Stage~2 free exploration \& reflection) analyzed via reflexive thematic analysis. Section~\ref{sec:s2_quant} presents the quantitative results, covering subjective data validity, objective extraction fidelity against a human-labeled gold set, and tool usability. The qualitative findings and thematic insights derived from participants' open-ended reflections are reported separately in Thematic Insights (Section~\ref{sec:results}).

\subsection{Methodology}
\label{sec:study2_method}

To evaluate the fidelity of our dataset and the usefulness of the Mission Analysis Dashboard, we conducted a three-stage mixed-methods study with players and designers. Participants were expected to connect their personal experience as players with their observations from the tool, providing meaningful feedback for design practice. Accordingly, we treat experienced players as experts-by-experience whose reflections on experiential dynamics complement those of professional designers; this purposive inclusion follows information power in qualitative sampling and lead-user reasoning, and helps surface fine-grained pacing/salience cues observed in game expertise research \cite{green2003action, malterud2016sample}. Study~2 comprises (i) quantitative validity and usability measurements (Stages~1 and~3) and (ii) qualitative reflection and reflexive thematic analysis (Stage~2; Section~\ref{sec:thematic_analysis}).

\subsubsection{\textbf{Participants}}
\label{sec:participants}

Participants were recruited via two methods. Experienced players were sourced through public calls on platforms such as Reddit (e.g., r/truegaming, r/videogames, r/SampleSize) and our university's gaming forums, followed by email-based screening to verify eligibility criteria. Designers were contacted directly via industry studios or game developer platforms (e.g., a formal game developer Discord channel) and invited through personalized invitations, with eligibility confirmed through brief online interviews. We analyzed data from 68 participants (50 male, 18 female; age 19-35, $M{=}24.5$, $SD{=}3.2$) spanning both player and designer roles (44.1\% experienced players, 27.9\% aspiring/indie designers, 19.1\% casual players, 8.8\% professional quest/mission designers). A full breakdown, including open-world familiarity, analytics-tool experience, and design-experience buckets, is provided in Appendix~\ref{app:demographics}.

We offered no monetary compensation. After providing informed consent, participants completed a self-paced online study (\textasciitilde30 minutes). Initially, participants filled out a brief demographic survey (2 minutes) capturing their background, game-design experience, tool familiarity (general game-analytics/visualization tools), and open-world games played. Next, they freely explored our web-based interactive tool to familiarize themselves with visualization results ($<30$ minutes) before proceeding through the three study stages described below.

\subsubsection{\textbf{Quantitative Method}}
\label{sec:quant_method}

The quantitative component consisted of Stage~1 (Data Validity Check) and Stage~3 (Usability \& Utility Evaluation). We designed Stage~1 to assess the accuracy and comprehensiveness of the automatically generated data artifacts. Participants selected 1-2 familiar games and rated Likert-scale items regarding action-list consistency and coverage, generated sequence completeness and correctness, and overall data credibility (Table~\ref{tab:stage1}). They also provided optional free-text responses highlighting missing or redundant actions, thus surfacing and quantifying extraction errors.

\begin{table}[tb]
  \caption{Stage 1 Items (Likert 1 = very strongly disagree … Likert 7 = very strongly agree).}
  \label{tab:stage1}
  \begin{tabular}{@{}lll@{}}
    \toprule
    Code & Prompt & Type \\ \midrule
    A1 & Action-List Consistency & Likert 7 \\
    A2 & Action-List Coverage & Likert 7 \\
    A3 & Missing Actions (optional list) & Open text \\
    A4 & Redundant Actions (optional list) & Open text \\
    A5 & LLM Sequence Completeness & Likert 7 \\
    A6 & LLM Sequence Correctness & Likert 7 \\
    A7 & Overall Data Credibility & Likert 7 \\ \bottomrule
  \end{tabular}
\end{table}

Stage~3 evaluated participants' experience with the tool itself through established quantitative measures: the System Usability Scale (SUS, 10 items, 5-point Likert, 0-100 score~\cite{brooke1996sus}), the short form User Experience Questionnaire (UEQ-S, 8 semantic-differential items rated from $-3$ to $+3$~\cite{hinderks2017design}), and the Single Ease Question (SEQ, single 7-point item rating overall ease~\cite{sauro2009comparison}). Additionally, participants provided open-ended feedback on the most helpful features, any confusion encountered, and areas for potential improvement.

All quantitative data were summarized consistently: Stage~1 Likert ratings with medians and inter-quartile ranges (IQR), and Stage~3 usability metrics (SUS, UEQ-S, SEQ) with both means and standard deviations (SD) as well as medians/IQR. Differences between players and designers were assessed via individual Mann-Whitney $U$ tests for each metric, with Holm correction for multiple comparisons. Effect sizes were reported as rank-biserial correlation ($r_U$) for Mann-Whitney tests, with 95\% bootstrap CIs; we interpret $r_U$ magnitudes as small (0.10-0.30), medium (0.30-0.50), and large ($>0.50$).

Ethical approval was obtained from our institutional review board. Participation was voluntary and anonymous; identifiable records are retained only until de-identification, which will be completed within 12 months of data collection or within three months after first publication, whichever occurs first. At that point the linkage key is destroyed. Before de-identification, participants could request withdrawal and their data would be deleted; after de-identification or publication, withdrawal is no longer possible. De-identified data are stored on an access-controlled institutional cloud for at least seven years from first publication, are not shared outside the lab, and any future reuse by the same team for related studies requires explicit participant opt-in.

\subsubsection{\textbf{Thematic Analysis}}
\label{sec:thematic_analysis}

Our qualitative corpus comes from Stage~2 (Free Exploration \& Reflection) of Study~2: participants used the dashboard to explore missions freely and responded to at least three guided reflection prompts covering cross-game commonalities, good versus poor mission patterns, mission/action evolution, and pacing strategies (Table~\ref{tab:stage2}; full text in Appendix~\ref{app:stage2-prompts}).

\begin{table}[tb]
  \caption{Stage 2 Reflection Prompts (Choose any $\geq3$). Full questions in Appendix~\ref{app:stage2-prompts}.}
  \label{tab:stage2}
  \begin{tabular}{@{}ll@{}}
    \toprule
    ID & Question (abridged) \\ \midrule
    Q1 & Six-dimensional commonalities/divergences across games \\
    Q2 & Good vs.\ flawed mission—dimensional centers/voids \\
    Q3 & Action diversity and block structure across games \\
    Q4 & Shared/missing block combos (good vs.\ poor missions) \\
    Q5 & Long-mission pacing differences \\
    Q6 & Neglected or over-emphasized dimensions/categories \\
    Q7 & Main vs.\ Side vs.\ POI mission priorities \\
    Q8 & Mission/action evolution across sequels \\
    Q9 & Formula-breaking uniqueness vs.\ monotony cases \\
    Q10 & Core \& flavoring ingredients of a good mission \\
    Q11 & Mapping dimensions to player motives (examples) \\
    Q12 & Peaks/valleys as emotional climaxes or recovery buffers \\ \bottomrule
  \end{tabular}
\end{table}

In accordance with reflexive thematic analysis as articulated by Braun and Clarke \cite{braun2006using, braun2019thematic, braun2019reflecting}, the first author, who had prior familiarity with most of the selected titles, led the analysis. For each title, the analyst responsible for coding and theme review was a coauthor with documented familiarity arising from prior play and/or analysis of developer materials and community walkthroughs. We treat prior exposure as contextual knowledge rather than evidence and mitigate bias through reflexivity memos, peer debriefs, and negative-case checks \cite{braun2006using, bowman2023using}. The TA corpus comprised the Stage~2 open-ended written reflections; we did not code wiki text, telemetry, or tool logs for TA. Analysts engaged with both gameplay (where applicable) and fandom discourse (e.g., walkthrough summaries), attending to semantic and latent patterns while avoiding premature, surface-level topic coding \cite{braun2019thematic}.

Following initial immersion, the authors inductively generated codes capturing meaningful language about mission structure, design reasoning, and gameplay reflections. Coding focused on user expressed summaries or design narratives rather than pre-coded categories, preserving interpretive richness. As codes accumulated, we clustered them into candidate themes reflecting recurring conceptual patterns in mission design (e.g., ``trade-off between narrative and combat intensity'' or ``repeated traversal-dialogue-combat block motifs''). Code clustering and initial theme delineation followed a reflexive iterative cycle, with theme definitions refined in joint analytic sessions \cite{vakeva2025don}.

We then reviewed candidate themes iteratively, verifying each theme's coherence, boundary, and prevalence in the corpus, merging overlapping themes and discarding those lacking analytic depth. This process continued until themes were internally consistent, conceptually coherent, and grounded in representative coded segments. Reflexivity memos aided transparency, documenting analytic decisions and researcher assumptions.

Through open, inductive coding, we first generated an initial set of candidate patterns, including (1) \textit{Dimensional Trade-offs}, (2) \textit{Action Block Grammar}, (3) \textit{Quest Category Variation}, (4) \textit{Cross-Game Design Patterns}, (5) \textit{Pacing-Driven Emotional Arcs}, (6) \textit{Novelty versus Redundancy}, and (7) \textit{Emergent Play}. Following Braun and Clarke's recursive phases of theme development \cite{braun2006using, braun2019thematic, braun2019reflecting}, we repeatedly reviewed, merged, and renamed these provisional themes, drawing on the twin criteria of \emph{information power} \cite{malterud2016sample} and \emph{conceptual depth} \cite{nelson2017using} to decide when further subdivision or consolidation was warranted. This iterative sense-making ultimately converged on four reportable themes (Section~\ref{sec:results}); the remaining three were merged or de-emphasized: (5) into (1) and (2), (6) into (1) and (4), and (7) reflected in quotations but not retained as a standalone theme given our focus on authored, critical-path flows (Section~\ref{sec:lim}).

By combining inductive coding with thematic clustering sensitive to our MAQV framework and visual analytics, our thematic analysis integrates qualitative depth with quantitative insight, bridging numerical patterns and experiential meaning.

\subsection{Quantitative Results}
\label{sec:s2_quant}

\begin{figure*}[t]
  \centering
  \includegraphics[width=0.95\linewidth]{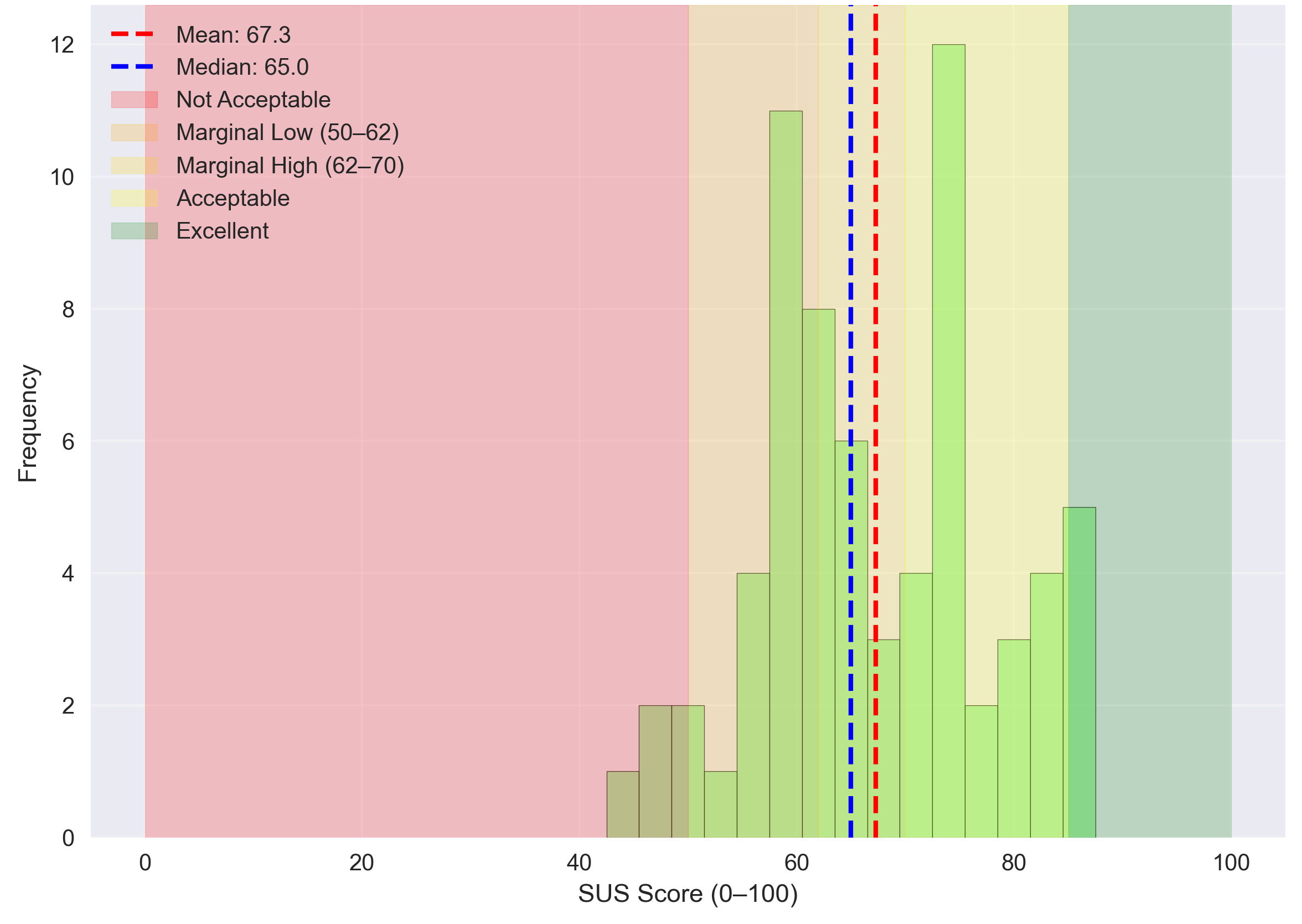}
  \caption{SUS score distribution with rating zones. Mean $=67.3$, Median $=65.0$; scores cluster around ``marginal-high'' with a tail into ``acceptable''.}
  \Description
  {SUS histogram with rating zones, where most scores in Marginal High and Acceptable zones. Histogram of SUS scores on the 0-100 scale with background bands: Not Acceptable (0-50), Marginal Low (50-62), Marginal High (62-70), Acceptable (70-85), Excellent ($\geq$85). Most responses fall between 60 and 80, straddling the Marginal High and Acceptable zones. Vertical dashed lines show the mean (67.3) in the Marginal High range and the median (65.0) at the upper end of Marginal Low.}
  \label{fig:sus-hist}
\end{figure*}

This section reports three strands of quantitative evidence from Study~2: subjective data validity (Stage~1), objective extraction fidelity against a human-labeled gold set (Validation Study, Section~\ref{sec:validation}), and tool usability (Stage~3). Table~\ref{tab:quantSummary} aggregates key descriptive statistics; Figures~\ref{fig:sus-hist} and \ref{fig:ueq-analysis} provide distributional overviews.

\begin{table}[t]
  \centering
  \caption{Summary statistics for Stage 1 (Validity) and Stage 3 (Usability).  
           SUS is reported on the canonical $0$-$100$ scale; UEQ-S scores range
           from $-3$ to $+3$. Stage~1 cells report IQR; Stage~3 cells report SD. 
           Per-item distributions are provided in Appendix~\ref{app:extended-plots}.}
  \label{tab:quantSummary}
  \begin{tabular}{lccc}
  \toprule
  & \textbf{Median} & \textbf{Mean} & \textbf{SD / IQR} \\ \midrule
    \multicolumn{4}{l}{\textit{Stage 1 — Validity Check (Likert 1-7)}}\\
    A1 (Action-List Consistency) & 5.00 & 4.99 & [4.75-6.00] \\
    A2 (Action-List Coverage) & 5.00 & 4.84 & [4.00-6.00] \\
    A5 (LLM Completeness) & 5.00 & 4.53 & [4.00-6.00] \\
    A6 (LLM Correctness) & 5.00 & 4.72 & [4.00-6.00] \\
    A7 (Data Credibility) & 5.00 & 5.01 & [4.00-6.00] \\ \midrule
    \multicolumn{4}{l}{\textit{Stage 3 — Usability \& Utility}}\\
    SUS score & 65.0 & 67.3 & 10.7 \\
    UEQ-S Pragmatic & 1.62 & 1.50 & 0.87 \\
    UEQ-S Hedonic & 1.25 & 1.16 & 0.80 \\
    SEQ (overall ease, 1-7) & 5.0 & 4.54 & 1.58 \\
    \bottomrule
  \end{tabular}
\end{table}

\paragraph{Data validity (Stage~1):}
Across the five validity items A1, A2, A5-A7, the median response is consistently 5/7, with means A1$=4.99$, A2$=4.84$, A5$=4.53$, A6$=4.72$, A7$=5.01$. Item-wise dispersion is small as reflected by IQRs (Q1-Q3) reported in Table~\ref{tab:quantSummary} (e.g., A1: 5.00 [4.75-6.00]; A6: 5.00 [4.00-6.00]; A7: 5.00 [4.00-6.00]). These results suggest agreement that the action lists and LLM-normalized sequences are adequate for the study's reflective analysis. Per-item distributions are provided in Appendix~\ref{app:extended-plots} (Figure~\ref{fig:stage1-dists}).

\paragraph{Extraction fidelity (Validation Study):}
Against the human-labeled gold set (80 missions; Section~\ref{sec:validation}), LLM predictions closely matched consensus labels, achieving micro F1 of 88.27\% (Precision 87.68\%, Recall 89.70\%), with 42.5\% exact sequence matches and 13.81\% normalized edit distance (95\% CIs via bootstrap). Performance was broadly consistent across mission types (Kruskal-Wallis, $p{=}0.3015$) and varied significantly across games ($p{<}0.001$), consistent with differences in mission complexity and vocabulary specificity (Table~\ref{tab:overall_f1}; Table~\ref{tab:type_f1}). The most frequent discrepancies were dialogue vs.\ cutscene substitutions, traversal granularity drift (merge/split), and occasional missing micro-steps.

\begin{table}[t]
  \centering
  \caption{Extraction performance vs.\ human gold (80 missions).}
  \label{tab:overall_f1}
  \begin{tabular}{lcc}
    \toprule
    Metric & Value & 95\% CI \\
    \midrule
    Precision $\uparrow$                    & 87.68\% & [82.14\%-93.22\%] \\
    Recall $\uparrow$                       & 89.70\% & [84.26\%-95.13\%] \\
    F1 (micro) $\uparrow$                   & 88.27\% & [82.87\%-93.66\%] \\
    Exact sequence match $\uparrow$         & 42.50\% & [31.43\%-53.57\%] \\
    Normalized edit distance $\downarrow$   & 13.81\% & [8.23\%-19.39\%] \\
    \bottomrule
  \end{tabular}
\end{table}

\begin{table}[t]
  \centering
  \caption{F1 by mission type in the validation study.}
  \label{tab:type_f1}
  \begin{tabular}{lcc}
    \toprule
    Type & $n$ & F1 (95\% CI) $\uparrow$ \\
    \midrule
    Main & 24 & 82.95\% \; [69.14\%-96.76\%] \\
    Side & 36 & 92.67\% \; [86.84\%-98.50\%] \\
    POI  & 20 & 86.72\% \; [75.98\%-97.47\%] \\
    \bottomrule
  \end{tabular}
\end{table}

\paragraph{Tool usability (Stage~3):}
The System Usability Scale shows \emph{borderline-positive} results (Mean$=67.3$, Median$=65.0$; Figure~\ref{fig:sus-hist}), placing the tool around the ``marginal-high'' band \cite{bangor2009determining} overall with a tail into the ``acceptable'' zone. We treat this as a baseline signal for an information-dense, coordinated multi-view research prototype rather than a definitive claim about polished usability; planned onboarding and interaction refinements are detailed in Section~\ref{sec:lim}. Following conventional SUS acceptability ranges (e.g., \cite{bangor2009determining,sauro2016quantifying,brooke1996sus}), a mean of 67.3 falls in the \emph{marginal-high} band and just below the commonly cited \emph{acceptable} threshold of 70, indicating borderline-acceptable usability with room for refinement. UEQ-S scores are likewise positive (pragmatic Mean$=1.50$, Median$=1.62$; hedonic Mean$=1.16$, Median$=1.25$) {\small[benchmark: pragmatic=\textit{Good}, hedonic=\textit{Above average} \cite{schrepp2015user, hinderks2018benchmark}]}, suggesting the interface is both practical and pleasant. Taken together, we regard the prototype as borderline-acceptable for first-use reflective tasks, with clear headroom for polish. The Single Ease Question (SEQ) indicates moderate ease (Mean$=4.54$, Median$=5.0$). Comparing experienced players and designers via two-sided Mann-Whitney tests with Holm correction found no significant differences after correction; for transparency we report Holm-adjusted $p$ and rank-biserial effect sizes $r_U$ (unadjusted $p$ in parentheses) as point estimates: Stage~1 Average $U{=}516.0$, $p_{\mathrm{adj}}{=}1.00$ $(p{=}0.4925)$, $r_U{=}{-}0.108$; SUS $U{=}431.0$, $p_{\mathrm{adj}}{=}1.00$ $(p{=}0.6411)$, $r_U{=}0.074$; UEQ-S Pragmatic $U{=}536.5$, $p_{\mathrm{adj}}{=}1.00$ $(p{=}0.3319)$, $r_U{=}{-}0.153$; UEQ-S Hedonic $U{=}486.5$, $p_{\mathrm{adj}}{=}1.00$ $(p{=}0.7782)$, $r_U{=}{-}0.045$; SEQ $U{=}483.0$, $p_{\mathrm{adj}}{=}0.81$ $(p{=}0.8116)$, $r_U{=}{-}0.038$. Effect sizes were small overall. Per-item SUS and UEQ-S histograms are shown in Appendix~\ref{app:extended-plots} (Figure~\ref{fig:sus-indiv} and \ref{fig:ueq-indiv}).

\paragraph{Implications:}
Together, the objective extraction fidelity (micro-F1 $=88.27\%$) and Stage~1 validity (medians $=5/7$ across A1-A7) indicate that our MAQV taxonomy and LLM pipeline yield trustworthy mission representations. Usability is borderline-acceptable (SUS mean $=67.3$, median $=65.0$); nevertheless, UEQ-S means (pragmatic $=1.50$, hedonic $=1.16$) and SEQ (Mean $=4.54$) are positive, supporting practical viability for first-use reflective tasks. We therefore treat the current system as analysis-ready while acknowledging headroom for polish; Section~\ref{sec:lim} outlines lightweight onboarding and clearer affordances to reduce initial cognitive load. These quantitative findings motivate and contextualize the qualitative thematic insights reported in Thematic Insights (Section~\ref{sec:results}).

\section{Thematic Insights}
\label{sec:results}

In this section, we categorize our qualitative exploration of open-world mission design into four thematic categories. These themes capture how designers and players perceive, navigate, and evaluate the interplay of mission dimensions and structures. The four themes we developed were: (1) Dimensional Trade-offs, (2) Action Block Grammar for Mission Flow, (3) Functional and Structural Variation by Quest Category, and (4) Cross-Game Design Patterns. Subsection headers correspond to codes from our initial coding pass; within each subtheme we unpack the distinct claims within that code and support them with representative quotations. Quotations are lightly edited to preserve voice and intent while normalizing slang, acronyms, and memes to common vocabulary, expanding shorthand, and omitting tangential fragments, retaining only core, thematically relevant content. While each theme is presented separately for clarity, the coding was interwoven: individual codes and quotations often connected to multiple themes, reflecting the multifaceted nature of open-world mission design practices and player engagement.

\subsection{Theme 1: Dimensional Trade-offs}

This theme examines how MAQV dimensions trade off to produce engaging missions: which dimensions act as pillars, when balanced baselines outperform skewed profiles, the two recurring archetypes, and how these patterns align with player motives.

\subsubsection{Foundational Pillars with Variable Depth}

Combat and Exploration are the structural foundations of most open-world missions, serving as the basic gameplay loop across genres and titles. Other dimensions, especially Narrative and Emotion, vary in presence and depth, giving each game or quest its distinct identity.

\begin{quote}
\textit{``Looking at the radar charts, you can really see how each game focuses on different things. In Spider-Man 2, everything feels balanced: lots of swinging around, fighting, and exploring, with the story and cutscenes tying it all together. GTA V, on the other hand, is all about the story moments and character stuff, but the gameplay itself is mostly driving and shooting, so it's not as strong on puzzles or unique mechanics. Far Cry 6 is heavy on action and exploration, but the story takes a back seat and you end up doing a lot of the same things. Honestly, most open-world games build around combat and exploration, but how much they focus on narrative or emotion is what really sets them apart.''}
\end{quote}

Narrative and Emotion dimensions act as complements rather than substitutes; when applied in the right dose they function as a ``universal tonic,'' and their indispensability grows with mission length. Narrative usually carries Emotion, whereas high Emotion alone does not guarantee a compelling Narrative arc. Emotion combined with Narrative consistently appears in missions participants explicitly praised, but continuous high-Emotion peaks can reduce player sensitivity to narrative twists—an ``emotional saturation'' effect.

\begin{quote}
\textit{``For me a well-designed mission contains high emotion and narrative engagement (Cyberpunk 2077, Who Wants to Live Forever); a flawed mission usually wastes time on exploration (The Ride). On the flip side, I've played missions that hit big emotional beats but barely moved the story forward, and they just didn't land the same way. The best ones combine emotion and narrative together so each heightens the other, rather than relying on one to carry the experience.''}
\end{quote}

Uniqueness is necessary but not sufficient; missions still fail when uniqueness is not balanced by other dimensions, and excess uniqueness cannot compensate for poor dimensional balance. Distinctive narrative content, rather than mechanical novelty alone, is the decisive factor in successful missions.

\begin{quote}
\textit{``I do feel uniqueness is really important for open world missions. I really love Spider-Man games, and I never feel bored when swinging through the city. But that is because the missions do not only cover these actions. Being a ninja in Rise of the Ronin is also cool (with a not-bad combat system). But I just want to finish the game as soon as possible after a lot of missions with only the ``Unique'' combat objectives.''}
\end{quote}

Puzzle-solving is almost always underemphasized or missing. Problem-Solving and Combat tend to suppress each other, while Exploration tolerates either. Missions overloaded with Problem-Solving are generally rated lower, and the lack of puzzles, investigations, or mini-games can make the pacing feel hollow. While many players report fatigue with puzzle-heavy segments, others explicitly ask for at least some puzzle content; this pattern raises an important but unresolved question beyond the evidence in our study: is the steady simplification and scarcity of puzzles in contemporary AAA titles driven mainly by studio-side cost-benefit and designer expertise constraints, or by a genuine decline in player appetite and demand for puzzle mechanics? Synthesizing our coded claims, we also observe that token, perfunctory insertion of a few puzzles, likely symptomatic of this trend, tends to alienate both constituencies.

\begin{quote}
\textit{``Most open-world games set you in a loop of combat and travel, with puzzles barely getting a look-in. Far Cry 6 is mostly shootouts and sneaking. It's fun for a bit, but without the investigation or mini-game, it starts to blur together. Spider-Man at least sprinkles in the occasional puzzle or stealth section, which helps break things up, but a lot of games skip that entirely. Sure, too many puzzles can drag the pace down, but having none at all leaves missions feeling like they're missing a layer.''}
\end{quote}

\subsubsection{Dimensional Balanced vs Specialized}

Missions read as strongest when a balanced baseline (no zeros across dimensions) is in place and one pillar is allowed to peak as a highlight. The peak is a bonus that turns a good, well-rounded quest into a memorable one. By contrast, single-axis missions without supporting beats (no story hook, no exploratory relief, no light problem-solving) tend to feel hollow—unless that lone axis reaches a truly exceptional bar, which is rare in open-world quest flow.

\begin{quote}
\textit{`` `Gangs of Novigrad' works because it has story, some exploring, a fight, a couple of light puzzles—and then it leans a bit harder into narrative. That extra push makes it stick. Compare that to a Far Cry outpost or a Horizon fetch run: it's basically travel + combat on repeat. Unless the setup is truly special, the missing story or combat parts make the whole thing feel weird.''}
\end{quote}

\subsubsection{Dimensional Archetypes}

Most open-world missions tend to fall into one of two broad types: narrative-driven or combat-driven. Narrative-oriented missions focus heavily on story, supported by emotion and exploration, while combat-oriented missions center on fighting, with exploration and emotion playing smaller supporting roles. These patterns give missions recognizable ``shapes'' (Figure~\ref{fig:mission-radar-per-game}) that set expectations and reflect the trade-offs designers make. While not all games conform rigidly to these archetypal shapes, the prevailing tendency is to expand upon the generic profile: adding breadth or height along its dominant axes, rather than to compress it further.

\begin{quote}
\textit{``When I looked across all the games on the radar charts, I kept seeing two clear shapes. One leans to the upper left, narrative first, then strong emotion and some exploration, kind of like a teardrop pointing to story. The other leans slightly to the upper right, peaking on combat with moderate exploration and emotion, but usually dropping off in puzzles and story. Simple, but it matches how I've always thought about these games.''}
\end{quote}

\subsubsection{Dimension-Motive Alignment}

Dimensions map directly to player motives: \textsc{Combat+Problem-Solving} feed challenge and mastery, \textsc{Narrative+Emotion} drive story and immersion, and \textsc{Exploration+Uniqueness} satisfy curiosity and novelty. Strong missions sequence these motives so each beat hits at least two at once—for example, opening with a narrative-emotion hook, giving space for exploration with unique mechanics, layering in a challenge that blends combat and puzzles, and ending with a payoff that ties back to the hook. Skipping an entire motive risks making the mission feel either hollow (all action, no reason) or static (all story, no play).

\begin{quote}
\textit{``The missions that stick with me do two things: give me a reason to care, then let me do something about it. Let me look around a bit, then make me plan a route and do related setup before the fight, and close with a payoff that ties back to the hook. If it's just map-markers and brawls, I zone out; if it's only cutscenes, I get restless. The sweet spot is when story, curiosity, and a small skill test all show up in the same run.''}
\end{quote}

\subsection{Theme 2: Action Block Grammar for Mission Flow}

We treat missions as chains of action blocks. In this theme, we analyze how sequencing, alternation, multi-phase scaffolding, and peak-valley rhythm shape moment-to-moment flow.

\subsubsection{Characteristic Action Blocks}

Across nearly every title, a small set of blocks forms the common backbone: \emph{combat} and \emph{traversal} (plus a franchise-specific ``special'' verb such as web-swinging, parkour, or scanning). By comparison, \emph{puzzles} and \emph{stealth} show up less consistently and are often underweighted. This bias produces familiar loops that feel good minute-to-minute but can blur together without occasional cognitive or tactical shifts.

\begin{quote}
\textit{``Most open-worlds have the same backbone: travel, then combat. In Far Cry 6, I drive or sprint in and it turns into a shootout; in Spider-Man 2, I swing to a marker and end up in an arena brawl; in Dying Light, it's parkour into a scrap. Puzzles and stealth only appear in a small number of titles. I am really sad to see that trend (Is stealth game really dead ?). But one thing is pretty interesting but still making sense: some games have their special actions at the top of their action diversity graph, like Dying Light.''}
\end{quote}

Every series tends to stitch these blocks into a recognizable signature: a recurring ``sentence'' of actions players learn to anticipate. These grammars set expectations and micro-pacing: a bit of movement, a quick look/handle moment, a small decision or puzzle, then a payoff; or a straightforward talk $\rightarrow$ go $\rightarrow$ fight template. The specifics differ by franchise, but the idea is the same: familiar sequences shape how a mission feels from the first step.

\begin{quote}
\textit{``In the Spider-Man games, the design formula is clear: swing across town, scan or pull a panel, solve a quick puzzle, then the fight kicks off. In medieval RPGs like Kingdom Come: Deliverance, Skyrim, or The Witcher 3, it's talk to someone, travel to the spot, and poke around the scene. Across a lot of series you see the same basic template—social setup, the trip, then the fight.''}
\end{quote}

\subsubsection{Alternation and Phase Structure}

Strong missions alternate between diverse action categories; weak ones loop the same strand without relief.
\begin{quote}
\textit{``In Far Cry 6 I kept falling into a shoot $\rightarrow$ drive $\rightarrow$ shoot loop from one outpost to the next. GTA mixes things up more (drive, then a dialogue setup, then drive, then a shootout or a mini-game, then a cutscene) so even when you revisit familiar verbs, the sequence feels less samey.''}
\end{quote}

High-quality missions move through multiple well-planned phases (setup, infiltration, execution, escape, resolution) rather than jumping straight to a single mode.
\begin{quote}
\textit{``A strong RDR2 tobacco-field stealth run flows stealth $\rightarrow$ planning $\rightarrow$ sabotage $\rightarrow$ escape $\rightarrow$ dialogue: stealth, destruction, combat, narrative. A weak Starfield ruin-scan is just flight $\rightarrow$ landing $\rightarrow$ scan $\rightarrow$ return: pure exploration with no combat or plot blocks, a flat trough on the radar.''}
\end{quote}

A planning/setup phase frames later mechanics and raises stakes; missions that skip it lose momentum and texture. By making the upcoming beats legible, the planning phase gives players a clear expectation scaffold that offsets early build-up fatigue; provided the later sequence pays off, the explicit promise of delayed gratification smooths the transition period and sustains engagement.

\begin{quote}
\textit{``The bank heist goes planning (dialogue/exposition; feels like another real side quest) $\rightarrow$ stealth infiltration $\rightarrow$ tense shoot-out $\rightarrow$ escape, different blocks that keep momentum. The debt-collection job is travel $\rightarrow$ quick fight $\rightarrow$ travel. Same traversal/combat bones, but without setup or escape phases the latter feels static next to the heist.''}
\end{quote}

\subsubsection{Climax and Recovery: Peak-Valley Grammar}

Healthy pacing alternates peaks (back-to-back combat or narrative-emotion spikes) with valleys (lighter traversal, dialogue, or small puzzles). Too many peaks exhaust; valleys that run long bore. Good arcs weave minor buffers even inside a single quest so players breathe before the next high.
\begin{quote}
\textit{``On the radar-timeline, stacked Combat/Emotion segments mark the big moments—bosses, reveals, set-pieces—while lower-key exploration or dialogue creates short breathers. The best flows alternate those peaks and breathers; when everything is a peak you burn out, and if the valley drags you tune out.''}
\end{quote}

\subsection{Theme 3: Functional and Structural Variation by Quest Category}

We examine how mission category (main, side, POI) shapes both function and structure: mains carry narrative load and context shifts, sides narrow mechanics and simplify flow, POIs deliver single-focus tasks, and long missions require multi-stage pacing with higher narrative support.

\subsubsection{Narrative Backbone of Main Quests}

Main missions typically push story and emotion while keeping exploration and combat steady; side quests pivot toward problem-solving and uniqueness; POIs are quick hits of exploration/combat with minimal story.

\begin{quote}
\textit{``In `The Witcher 3', it's pretty clear: mains are heavy on story and feelings, moving between talk, clues, travel and a big fight. Side quests/contracts are more about prep and puzzles—follow a trail, oil a blade, plan a takedown. POIs are the opposite: ride in, clear a camp, loot, leave. Each type has a different job.''}
\end{quote}

\begin{quote}
\textit{``Main missions are mostly cinematic, emotionally invested and layered while side missions feel more experimental and world-building (sometimes morally interesting). POIs are usually mechanically repetitive and have relatively lower narrative investment.''}
\end{quote}

When a main quest swaps the usual context or role, it breaks routine and sticks in memory.

\begin{quote}
\textit{``RDR2's `A Fine Night of Debauchery' flips the script into a high-society poker heist with a tense getaway—barely any straight gunfights. Horizon Forbidden West's relic ruins do a similar thing with self-contained puzzle spaces.''}
\end{quote}

Big, designed turns like twists, bespoke mechanics, and branching create the peaks people remember; template errands fade fast.

\begin{quote}
\textit{``Quests like `A Night to Remember' in W3 or Cyberpunk's Heist intro land because they bend the formula—branching choices, unique mechanics, big reveals. Routine fetches or basic bounties feel like background noise next to those set-pieces.''}
\end{quote}

\subsubsection{Dimensional Bias}

Main quests usually keep multiple dimensions in play with more complex, interleaved flows, while side quests become simpler and more linear and POIs are one-and-done.

\begin{quote}
\textit{``In a lot of games, mains have flower-like pattern in all dimensions: story, combat, exploration, even a bit of puzzle all show up. Sides trim down to a couple of beats, and POIs are tiny one-shots. The complexity scales with narrative importance, and you can feel it in the flow.''}
\end{quote}

Side missions tend to narrow the mechanic set, and POIs spike a single axis (often exploration or a gimmick) while keeping everything else low.

\begin{quote}
\textit{``Main quests pair strong story with a few puzzles; sides keep some story but cut the big combat set-pieces; POIs are quick hits—explore a spot, do one simple interaction, move on. They help pacing, but you shouldn't expect the full range there.''}
\end{quote}

\subsubsection{Long Mission Design}

Good long missions keep a centered profile and rotate verbs to manage fatigue; weak ones tunnel on a single axis and drag.

\begin{quote}
\textit{`` `Deep Secrets of the Earth' or `Alduin's Wall' stay balanced in sneak, travel, puzzle, fight, cutscene, so the pace breathes. Something like `A Kind and Benevolent Despot' leans hard on emotion+combat for ages with little reset; no puzzle or exploration beats to break it up, so it starts to feel long.''}
\end{quote}

Strong long quests shift the spotlight, so the curve rises and falls instead of keeping static.

\begin{quote}
\textit{``Good long missions shift focus: find some clues, talk it through, fight, then make a choice that matters. When a mission sits on one activity, like waves of enemies or pure fetching, the curve goes flat and I just tire out.''}
\end{quote}

Epic finales work as distinct acts with buffers; monotone marathons don't.

\begin{quote}
\textit{``Think `Battle of Kaer Morhen' or Tsushima's castle assault: prep, short breath, defense, duels, payoff. Travel - talk - defend - puzzle - fight in waves. Compare that to an endless bandit-camp loop—travel, fight, loot, repeat—and you see why one soars and the other grinds.''}
\end{quote}

The longer a mission runs, the more story scaffolding it needs to justify the time and mix up the beats.
\begin{quote}

\textit{``If a quest runs an hour, it can't just be a long corridor of fights. Give me why we're here, a twist halfway, a choice that matters, and a payoff that ties back. The longer it goes, the more narrative fuel it needs or the action starts to feel like padding.''}
\end{quote}

\subsection{Theme 4: Cross-Game Design Patterns}

This subsection examines cross-game patterns: dimensional ``fingerprints,'' studio/genre clusters, shared block skeletons versus layering and pacing, and how series evolve across sequels.

\subsubsection{Dimensional Analysis Across Games}

Across titles, a small set of dominant dimension pairs reliably define identity: clear peaks shape player expectations long before the first fight or cutscene.

\begin{quote}
\textit{``On the radars every game has a tell. RDR2 leans hard into exploration+emotion (long rides, quiet moments), Witcher 3 spikes on narrative+emotion with steady combat, and Tsushima pushes combat+uniqueness (stealth duels, swordplay). Different mixes, instantly recognizable.''}
\end{quote}

At the catalogue level, games cluster into ``families'': studio habits and genre conventions pull dimension weights into similar regions, making those family traits easy to spot. This pattern implies that studios, having internalized these dimensional centers of gravity, often bias new productions toward established or adjacent profiles, intentionally referencing prior entries or comparable titles and thereby reinforcing the observed clustering.

\begin{quote}
\textit{``Bethesda games (Skyrim, Fallout) bunch near exploration + social-interaction + problem-solving; they're about roaming and figuring things out. Rise of the Ronin and Ghost of Tsushima both sit deep in the ninja+combat lane. Their dimension peaks and even mission action sequences line up almost perfectly, built around stealth approaches, duels, and parry-heavy swordplay fights.''}
\end{quote}

Signature mechanics tend to resonate with the world and theme; when they do, loops feel fresh, and when they don't, or get overused, formula fatigue sets in.

\begin{quote}
\textit{``Spider-Man stays fresh because swinging and little puzzles fit the fantasy; Far Cry's outpost loop, run too many times, goes boring. Tsushima's haiku and duels feel right for the setting; checkpoint grinds don't.''}
\end{quote}

\subsubsection{Action Block Structural Similarity}

Across franchises, missions often share a common skeleton (travel $\rightarrow$ encounter $\rightarrow$ narrative beat). What separates a great mission is how it layers and alternates blocks: dropping in stealth, investigation, or puzzles, creating peaks and breathers, rather than looping a single block to exhaustion.

\begin{quote}
\textit{``They share the same bones, but the craft feels different. The Spider-Man finale keeps swapping gears: swing, clear debris puzzles, sneak past nests, multi-phase boss, then an emotional scene, so it breathes. A mission like `Blood Ties' in Far Cry is basically checkpoint $\rightarrow$ gunfight on repeat; short cutscenes, not much to figure out... I got tired.''}

\end{quote}

\subsubsection{Series Analysis}

Across sequels, structural complexity usually grows: more staged missions, richer interlocks between systems, and bigger set-pieces—but size alone doesn't guarantee better play.

\begin{quote}
\textit{``Newer entries usually add phases and toys—heists get extra steps, finales come in acts, systems talk to each other more. It feels bigger and busier, but it only matters if those pieces actually change how you play.''}
\end{quote}

Sequels either widen the mechanic set (good) or scale up the same loop (fatigue). Added tools diversify action blocks; unchanged side loops keep the grind intact.

\begin{quote}
\textit{``Forbidden West's glider/grapple and trickier puzzles make more routes and block combos, but if collectibles and outposts run the same script, the bloat catches up. New verbs help; recycled checklists don't.''}
\end{quote}

Studios also shift design focus across entries—trading spectacle for story, or breadth for speed—creating different pacing and payoffs.

\begin{quote}
\textit{``RDR2 leans away from pure spectacle into slower, heavier narrative beats. CDPR's jump from Witcher 3 to Cyberpunk flips the feel—vertical city, faster combat flow—but some of the big, layered side-quest texture gets thinner.''}
\end{quote}
\section{Discussion}

Our thematic analysis, informed by the reflections of experienced players and designers interacting with our visualization tool, offers specific insights into how modern open-world missions are constructed and experienced. The study's findings introduce precise terminology that designers can use to break down mission structures, and they frame a perspective on where the open-world genre is heading with its current design formulas. In this section, we first reflect on the systemic design challenges our findings illuminate within the broader context of AAA game development. We then synthesize our themes into a set of actionable design heuristics aimed at fostering diverse mission experiences in Section~\ref{sec:toolkit}. From a visual analytics standpoint, the takeaway is a task-oriented coordination of familiar views that lets analysts move from portfolio-level emphasis and similarity to concrete sequencing for inspection and discussion. We present these conclusions as evidence that the tool is useful for reflection, rather than as a definitive validation of its effectiveness. Finally, we critically evaluate our own analytical framework, discussing its strengths and limitations as a reflective tool for designers and researchers.

\subsection{A Critical Reflection on Contemporary Open-World Design}

A significant critique of modern open-world games is their tendency towards ``checklist''-style design, where vast worlds are populated with repetitive, low-value content \cite{wang2024meaningful, nelson2023claustrophobia}. Our findings provide empirical weight to this critique, revealing a systemic imbalance in the experiential dimensions offered to players. Across numerous titles, participants consistently identified an over-emphasis on the dimensions of \emph{Combat} and \emph{Exploration} at the expense of \emph{Problem-Solving}, \emph{Uniqueness}, and often, deep \emph{Emotional} engagement. As one designer noted: across all four titles, Problem-Solving is systematically under-represented or even missing, while Exploration is often over-emphasized. 

This dimensional imbalance is not accidental but appears to be a product of the economic realities of AAA development. To justify ballooning production costs, major studios must target the widest possible audience \cite{kerr2017global}. Combat and exploration are mechanically straightforward, visually spectacular, and broadly appealing pillars. In contrast, complex puzzle design is a high-risk endeavor; it is difficult to balance, can frustrate less patient players, and offers a lower return on graphical investment compared to explosive set-pieces \cite{fullerton2024game}. The result is a repeatable, risk-averse core loop (``travel-fight''/``shoot-drive-shoot''). It succeeds commercially but produces uneven pacing and thin connective content, a pattern noted by players and designers in our study. While this formula has defined a generation of commercially successful games, it also risks creative stagnation, leaving significant potential for richer, more intellectually and emotionally varied experiences untapped. These observations motivate a synthesis on how ``formula'' should be treated in mainstream open-world development (see Section~\ref{sec:rethink}).

\subsection{Rethinking the Role of Design Formulas}
\label{sec:rethink}

From our tool-assisted prompts and reflections, a broad consensus emerged: design formulas do exist, and many ideas and patterns are shared across modern open-world games. Synthesizing Themes 1-4 and participant reflections, we use ``formula'' descriptively rather than prescriptively. Users repeatedly observed that even the most celebrated titles reveal shared structures in our analysis, including clear inheritance within a series and recognizable overlaps across studios. Yet those same games feel distinct because the ordering, emphasis, and pacing of similar ingredients are tuned to their verbs, theme, and fiction, yielding different moment-to-moment experiences from ostensibly similar scaffolds. The pathology appears when studios copy-paste the same design formula across different games and overuse it within a single title, producing large volumes of undifferentiated, non-targeted tasks. Within mainstream, production-oriented open-world design, we therefore treat formulas as diagnostic baselines and a safe lower bound for analysis and coordination, but not templates to replicate. In that role, pattern libraries and style guides can help teams inspect and when desirable subvert defaults, guard against overlong single-axis plateaus, and surface under-explored combinations, so that cost and coherence improve without constraining distinctive, memorable play. This stance directly motivates the four actionable heuristics we articulate next in Section~\ref{sec:toolkit}.

\subsection{The Designer's Toolkit: A Framework for Crafting Memorable Missions}
\label{sec:toolkit}

While our analysis highlights systemic issues, it also deconstructs the mission ``formula'' into its constituent parts, offering a concrete toolkit for designing more varied and impactful missions. By moving beyond high-level narrative planning to measurable dimensional balance and explicit block sequencing, our findings can be distilled into a set of actionable heuristics. Concretely, we articulate four actionable heuristics (H1-H4): \textbf{H1} balanced baseline with selective emphasis; \textbf{H2} peak-valley pacing via action blocks; \textbf{H3} tier-appropriate roles across main/side/POI; and \textbf{H4} verb-first innovation rather than noun accretion. 

\textbf{H1:} First, the results indicate that memorable missions are typically characterized not by a single dominant dimension but by a deliberate balance across MAQV, followed by selective emphasis. Participants described this as a ``balanced star'' on the radar chart: establish a balanced baseline to avoid experiential gaps and then allow one or two dimensions to peak to create a distinctive identity. In concrete terms, designers can purposefully combine, for example, \emph{Exploration} with \emph{Problem-Solving} and \emph{Narrative} to jointly address curiosity, mastery, and immersion, rather than defaulting to a combat-exploration template.

\textbf{H2:} Second, pacing can be operationalized through an action block ``peak-valley'' grammar. The most engaging missions interleave high-intensity peaks (e.g., \emph{Combat} segments or salient \emph{Emotional} beats) with low-intensity valleys or recovery buffers (e.g., \emph{Exploration}, \emph{Problem-Solving}, or social interaction blocks), producing a sustainable rhythm of tension and release \cite{sweetser2005gameflow}. Conversely, extended single-activity plateaus (e.g., continuous combat waves) or overlong valleys depress engagement. The action block representation makes such alternation explicit and thus designable.

\textbf{H3:} Third, we observe consistent role differences by content tier. \emph{Main quests} are typically complex, multi-dimensional set-pieces that carry narrative and emotional weight; \emph{side quests} focus more on world-building or mechanical depth; and \emph{points of interest (POIs)} function as brief, single-focus encounters that punctuate exploration. Treating these categories as distinct roles within a coherent portfolio, rather than as interchangeable ``filler'', helps ensure that content at all scales contributes meaningfully to the overall experience.

\textbf{H4:} Finally, cross-franchise patterns indicate that durable innovation is driven less by adding new ``nouns'' (content or tools) to existing templates and more by expanding the player's core ``verb'' set and integrating those verbs into mission structure. Examples include the introduction of new traversal mechanics in \emph{Horizon Forbidden West} or expanded stealth options in \emph{Marvel's Spider-Man 2}, which, when embedded in the action block grammar, produce richer interaction, alter pacing, and diversify dimensional profiles. In practice, teams should prioritize verb-level additions and ensure missions require their use, rather than scaling world size alone.

During development, teams can operationalize H1-H4 using our mission-analysis pipeline (CSI $\rightarrow$ MAQV profiling $\rightarrow$ action-block extraction), which can consume stepwise inputs from quest scripts/blockouts (via CSI), QA step lists, or early playtest telemetry to support iteration.

\subsection{Limitations, Future Directions, and Reflections on Quantifying Mission Design}
\label{sec:lim}

We present MAQV and its visual analytics as a \emph{reflection probe} rather than an objective score of mission ``quality.'' The framework externalizes designers' and players' tacit judgments by making action block structure and experiential balance visible, offering a shared language to reason about pacing, emphasis, and formula. Participants reported that the MAQV quality-flow charts and block timelines helped them articulate why some missions felt repetitive while others felt planned and varied. Our aim, therefore, is not to judge taste, but to support disciplined reflection with personal experience and analyzable artifacts.

At the same time, several limitations bound our claims. First, \emph{dimensional reduction}: collapsing rich, situated play into six continuous dimensions introduces simplifications. The dimensions foreground experiential roles but do not capture aspects such as encounter design, timing and staging, camera work, or spatial audio, which also shape mission feel. Second, \emph{data provenance}: we rely on community-authored walkthroughs and LLM-based conversion to canonical action blocks. While our validation indicates good fidelity, the source texts vary in granularity and emphasis, and extraction errors or omissions can propagate. Third, \emph{coverage}: the pipeline privileges authored, critical-path flows. Optional beats, failure loops, and emergent play are only partially reflected, and branching is often linearized for analysis, which can understate variability. Fourth, \emph{comparability}: game-specific action vocabularies, smoothing choices, and normalization can influence cross-title contrasts, and aggregations (e.g., mission-level centroids) may hide within-game variance. Fifth, \emph{usability}: the information-dense, multi-view dashboard introduces unfamiliar constructs, so first-time use has a learning curve (SUS mean 67.3). Sixth, \emph{scope and selection bias}: because inclusion required substantial stepwise public walkthroughs, our sample skews toward high-visibility, conventionally structured open-world titles. We view this as a pragmatic and acceptable trade-off to guarantee dependable extraction and fair cross-title analysis, and we bound our claims accordingly. Seventh, \emph{priming}: our survey and task materials repeatedly referenced ``formulas,'' which may have biased participants toward judging how well a formula was executed rather than whether a formula should exist. This framing could inflate neutral or positive attitudes toward well-used formulas and should be considered when interpreting responses. Finally, \emph{visualization scope and encoding}: our visualization contribution is organizational rather than algorithmic: a coordinated arrangement of familiar views used as a reflection probe; we did not conduct a dedicated visualization study. The current charts prioritize readability over precision; radar overlays, PCA projections, and frequency plots are diagnostic cues, not absolute claims, and may be misread if treated as such.

These constraints suggest concrete improvements. On the UX side, we will add lightweight onboarding (guided tour and inline tooltips with worked examples), clearer legends/filters, and progressive-disclosure workflows in Compare mode (search and saved views), with the goal of improving first-use usability (higher SUS) without reducing analytic power. For example, we are prototyping a Gantt-style rendering of the action timeline (Fig.~\ref{fig:action-timeline}) to improve readability, expose relative durations, and support quick cross-mission alignment. Integrating engine or telemetry traces will let us triangulate designed flow against enacted play. This enables time-weighted MAQV timelines, failure-aware pacing profiles, and confidence intervals around dimension estimates. A designer-in-the-loop plug-in for common engines could support editable action sequences, ``what-if'' pacing experiments, automatic peak-valley detection, and safety checks (e.g., alerts for overlong single-axis plateaus). On the representation side, we will model missions as branch-preserving DAGs (nodes = action blocks; edges weighted by transition frequency/duration), enabling path alignment, subgraph motif mining, and branch centrality alongside linear timelines. Genre-aware weighting and adaptive dimension sets can improve construct validity across franchises. Methodologically, mixed annotation strategies that combine expert consensus with lightweight crowdsourced checks could reduce extraction bias; uncertainty visualization and per-mission provenance summaries would make quality transparent at the point of use. To address potential priming, we will vary prompt language and include non-priming control conditions, counterbalance task order, add manipulation checks, and preregister analyses to quantify and isolate framing effects. Upon acceptance, we will release a reproducible corpus (derived annotations only), analysis scripts, and dashboard-generated design worksheets to support longitudinal and cross-studio comparisons.

Stepping back, our emphasis is on quantifying mission structure and building visualizations for sense-making, not on prescribing a single formula or reducing judgment to a score. MAQV and the action block grammar are intended to complement narrative craft and encounter design by providing an analyzable, communicable structure. We see the framework as a supplementary tool for reflection. Used with designer expertise, player research, and production constraints, it helps teams reason about pacing, variation, and portfolio balance more clearly and with less reliance on tacit judgment.

\section{Conclusion}

Open-world missions are often criticized for repetitive, formulaic structures that drive player fatigue and content hollowness. To address this, we introduce a framework that deconstructs missions into quantifiable action blocks and experiential dimensions (MAQV). Using an interactive visualization tool as a reflection probe with expert players and designers, we conducted a thematic analysis that identified a practical grammar for mission design. Preliminary validation indicates that the extraction and scoring are sufficiently reliable for reflective analysis, and the tool shows acceptable first-use usability. Our findings identify three recurrent practices: deliberate balancing across MAQV dimensions; rhythmic sequencing of action blocks into peak-valley structures; and functional stratification of missions within a coherent content portfolio. Together, these results provide an evidence-based method and reusable tools, including representation, measures, and visualization, for day-to-day review. In practice, the framework enables teams to audit dimensional balance, pacing, and portfolio roles to produce varied, deliberately paced missions with clearer experiential intent. We offer this work as a first step toward rigorous, quantitative modeling of in-game player experience to inform design decisions and evaluation across production, and we hope it catalyzes broader efforts to integrate player-experience modeling into HCI and game development practice.

\begin{acks}
We thank our study participants for their time and thoughtful feedback.
This research is supported by NSERC Discovery Grant RGPIN-2019-05213.
\end{acks}

\bibliographystyle{ACM-Reference-Format}
\bibliography{sample-base}

\appendix
\section{Dataset Pipeline and Compliance Details}
\label{app:data}

This appendix details the dataset pipeline and compliance boundaries. We used a semi-automated Python pipeline to access pages via the MediaWiki Action API with polite rate limiting. For each mission name, the scraper generated candidate URL slugs to accommodate Fandom idiosyncrasies by normalizing spaces to underscores, testing common disambiguation suffixes (e.g., \textit{\_(Quest)}), and permuting case for articles and prepositions (e.g., “The\_of\_the” vs. “the\_of\_the”). The fetcher followed redirects and retried on transient errors. A heuristic parser prioritized textual sections such as \textit{Walkthrough}, \textit{Quest Stages}, or \textit{Description}, with a fallback to introductory paragraphs, while stripping navigation boxes, infoboxes, and advertisements.

Quality control retained only descriptions of at least 35 words, discarding stubs and empty pages. For each page, we recorded stable snapshot metadata—page \textit{revision\_id}, fixed snapshot URL with \textit{oldid}, ISO-8601 retrieval timestamp, license string (CC BY–SA), and a SHA-256 checksum of the raw HTML—and all analyses in the paper are computed on these immutable snapshots. Using wiki-curated lists, each mission was categorized as \textit{Main Quest}, \textit{Side Quest}, or \textit{Point of Interest (POI)}. The pipeline produced 2435 raw mission descriptions; after conversion to action sequences and validity filtering, the final dataset includes 2191 missions (607 Main, 1174 Side, 410 POI).

Compliance and redistribution boundaries are as follows. Except where otherwise noted on individual pages, Fandom text is licensed under CC BY–SA 3.0. We accessed content exclusively via the MediaWiki API in line with the site’s Terms of Use and processed text only; media assets (images, videos) were excluded due to heterogeneous licenses. For redistribution, we will release only derived, non-verbatim annotations—normalized action sequences and per-mission MAQV vectors—with per-page attribution links; we do not redistribute full wiki text. The corpus contains no personal data and was reviewed under our IRB protocol.

\lstdefinestyle{acmcode}{
  basicstyle=\ttfamily\footnotesize,
  numbers=left,
  numberstyle=\scriptsize,
  stepnumber=1,
  numbersep=6pt,
  frame=tb,
  rulecolor=\color{black!20},
  framerule=0.5pt,
  columns=fullflexible,
  keepspaces=true,
  showstringspaces=false,
  tabsize=2,
  breaklines=true,
  captionpos=b
}

\lstdefinelanguage{JSON}{
  basicstyle=\ttfamily\footnotesize,
  showstringspaces=false,
  morestring=[b]",
  morecomment=[l]{//},
  literate=
   *{:}{{{\color{black}{:}}}}{1}
    {,}{{{\color{black}{,}}}}{1}
    {\{}{{{\color{black}{\{}}}}{1}
    {\}}{{{\color{black}{\}}}}}{1}
    {[}{{{\color{black}{[}}}}{1}
    {]}{{{\color{black}{]}}}}{1}
}

\section{LLM Prompt Specification}\label{app:llm_prompt}

This appendix provides the full prompt specification used in the action‐block extraction pipeline (see Section~\ref{sec:actionblock}) and is intended to support transparency and reproducibility expected for LLM-enabled CHI research. Validation of extraction fidelity against a human-labeled gold set is reported in Section~\ref{sec:validation}.

\subsection{Prompt Structure}

The prompt comprises four components aligned with our extraction procedure (Section~\ref{sec:actionblock}):
\begin{enumerate}
    \item \textbf{System Message}: Declares the model’s role and constraints.
    \item \textbf{Action Library}: Game‐specific predefined actions.
    \item \textbf{Instruction Set}: Stepwise mapping rules.
    \item \textbf{Mission Description}: The input text to convert.
\end{enumerate}

\subsection{Complete Prompt Template}

\begin{lstlisting}[style=acmcode, language=Python, caption={Prompt construction function (Python pseudocode)}, label={lst:prompt_build}]
from typing import List
import json

def build_prompt(actions: List[str], desc: str) -> list[dict]:
    system = (
        "You are an expert game analyst who converts free-form mission "
        "descriptions into ordered lists of predefined game actions. "
        "Your task is to break down mission walkthroughs into elementary "
        "action steps and map each step to exactly one predefined action."
    )
    user = (
        "Predefined actions (choose only from this list):\n"
        f"{json.dumps(actions, ensure_ascii=False)}\n\n"
        "Instruction:\n"
        "1. Read the mission description.\n"
        "2. Break it into elementary action steps.\n"
        "3. Map each step to exactly one predefined action.\n"
        "4. Ignore narrative or non-action details.\n"
        "5. Output ONLY the ordered list of action names as a JSON array.\n\n"
        f"Mission description:\n\"{desc}\""
    )
    return [{"role": "system", "content": system},
            {"role": "user",   "content": user}]
\end{lstlisting}

\subsection{API Configuration}

\begin{lstlisting}[style=acmcode, language=Python, caption={LLM/API configuration used in our pipeline}, label={lst:api_config}]
MODEL            = "gpt-4.1"
REQUEST_TIMEOUT  = 30            # seconds
MAX_RETRIES      = 5
RATE_LIMIT_SEC   = 5
TEMPERATURE      = 0             # deterministic output
\end{lstlisting}

\subsection{Example Prompt Instance}

\begin{lstlisting}[style=acmcode, language=JSON, caption={Example prompt (Fallout 4 mission)}, label={lst:example_prompt}]
{
  "role": "system",
  "content":      
        "You are an expert game analyst who converts free-form mission descriptions into ordered lists of predefined game actions. Your task is to break down mission walkthroughs into elementary action steps and map each step to exactly one predefined action."
}
{
  "role": "user",
  "content": "Predefined actions (choose only from this list):\n[
    \"Walk / Sprint\",
    \"Dialogue Choice\",
    \"Speech Charisma Check\",
    \"Pip-Boy Fast-Travel\",
    \"Loot Container\",
    \"Workshop Build Object\",
    \"Stealth Boy Cloak\",
    \"Sneak Crouch\",
    \"Suppressed Shot\",
    \"VATS Target Shot\",
    \"Melee Bash\",
    \"Power-Armor Melee Slam\",
    \"Grenade Throw\",
    \"Vertibird Airdrop Ride\"
  ]\n\nInstruction:\n1. Read the mission description.\n2. Break it into elementary action steps.\n3. Map each step to exactly one predefined action.\n4. Ignore narrative or non-action details.\n5. Output ONLY the ordered list of action names as a JSON array.\n\nMission description:\n\"Mama Murphy insists that she can see the future ...\""
}
\end{lstlisting}

\subsection{Expected Output Format}

\begin{lstlisting}[style=acmcode, language=JSON, caption={Expected JSON output format}, label={lst:output_format}]
[
  "Dialogue Choice",
  "Speech Charisma Check",
  "Pip-Boy Fast-Travel",
  "Loot Container",
  "Workshop Build Object",
  "Dialogue Choice",
  "Speech Charisma Check"
]
\end{lstlisting}

\subsection{Error Handling and Validation}

\begin{lstlisting}[style=acmcode, language=Python, caption={Retry and output validation in the extraction pipeline}, label={lst:error_handling}]
def get_actions(actions: list[str], desc: str) -> list[str] | None:
    for attempt in range(1, MAX_RETRIES + 1):
        try:
            return call_chatgpt(actions, desc)
        except Exception as e:
            print(f" attempt {attempt}/{MAX_RETRIES} failed: {e}")
            if attempt < MAX_RETRIES:
                time.sleep(RATE_LIMIT_SEC)
    return None

def call_chatgpt(actions: list[str], desc: str) -> list[str]:
    resp = openai.ChatCompletion.create(
        model=MODEL,
        messages=build_prompt(actions, desc),
        temperature=TEMPERATURE,
        request_timeout=REQUEST_TIMEOUT
    )
    # Validate JSON array of strings
    arr = json.loads(resp.choices[0].message.content.strip())
    if not isinstance(arr, list) or not all(isinstance(a, str) for a in arr):
        raise ValueError("Model returned invalid JSON array.")
    return arr
\end{lstlisting}

\subsection{Action Library Structure}

Predefined actions are stored in JSON with the following fields:
\begin{itemize}
  \item \textbf{name}: canonical action identifier used in prompts;
  \item \textbf{scores}: MAQV scores (uniqueness, combat, narrative, exploration, problem\-/solving, emotional);
  \item \textbf{description}: human\-readable explanation;
  \item \textbf{category}: e.g., Traversal, Combat, Social Interaction.
\end{itemize}

\subsection{Reproducibility Notes}

To reproduce our results:
\begin{enumerate}
  \item Use the prompt composition in Listing~\ref{lst:prompt_build};
  \item Set \texttt{TEMPERATURE = 0} for deterministic outputs;
  \item Use the same action library schema (Section~\ref{sec:actionblock});
  \item Implement retry/validation as in Listing~\ref{lst:error_handling};
  \item Note minor variability may occur due to provider model updates.
\end{enumerate}

\subsection{Model Version Information}

\begin{itemize}
  \item \textbf{Primary Model}: GPT-4.1.
  \item \textbf{API}: OpenAI ChatCompletion.
  \item \textbf{Token limits/context}: All missions fit within the model context with our batching.
\end{itemize}

This specification complements our Method (Section~\ref{sec:actionblock}) and Validation Study (Section~\ref{sec:validation}), enabling other researchers to replicate the extraction pipeline on comparable corpora.

\section{Visualization Gallery and Implementation Notes}
\label{app:vis}

\begin{figure*}[t]
  \centering
  \includegraphics[width=\linewidth]{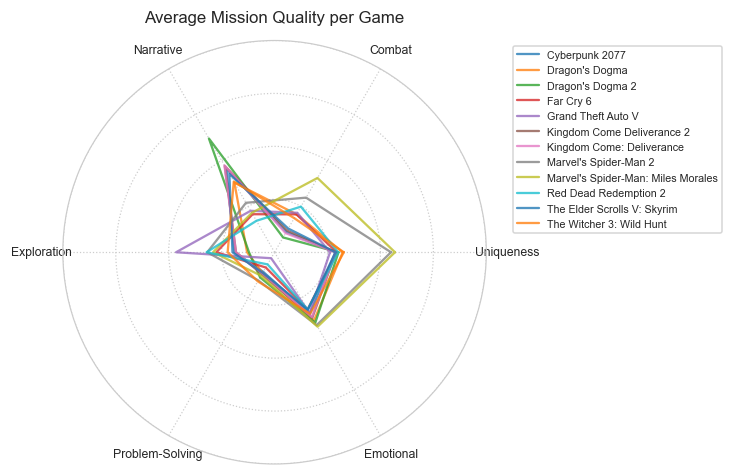}
  \caption{Combined radar chart comparing mission-dimension emphasis across selected games.}
  \Description{Radar chart titled Average Mission Quality per Game. The chart has six axes: Uniqueness, Combat, Narrative, Exploration, Problem-Solving, and Emotional. Each axis ranges from the center outward, representing higher average values for that dimension. Twelve open-world games are plotted, each shown with a colored line: Cyberpunk 2077, Dragon's Dogma, Dragon's Dogma 2, Far Cry 6, Grand Theft Auto V, Kingdom Come Deliverance, Kingdom Come Deliverance 2, Marvel's Spider-Man 2, Marvel's Spider-Man: Miles Morales, Red Dead Redemption 2, The Elder Scrolls V: Skyrim, and The Witcher 3: Wild Hunt. Most games cluster near the center for Problem-Solving and Emotional dimensions, showing low scores. Several games, such as Dragon's Dogma 2, extend further on the Narrative axis, while Marvel's Spider-Man games show stronger Uniqueness and Combat emphasis. Exploration values vary moderately across titles, with Grand Theft Auto V extending more than others. The visualization enables cross-game comparison of how mission design emphasizes different quality dimensions.}

  \label{fig:mission-radar}
\end{figure*}

\begin{figure*}[t]
  \centering
  \includegraphics[width=\linewidth]{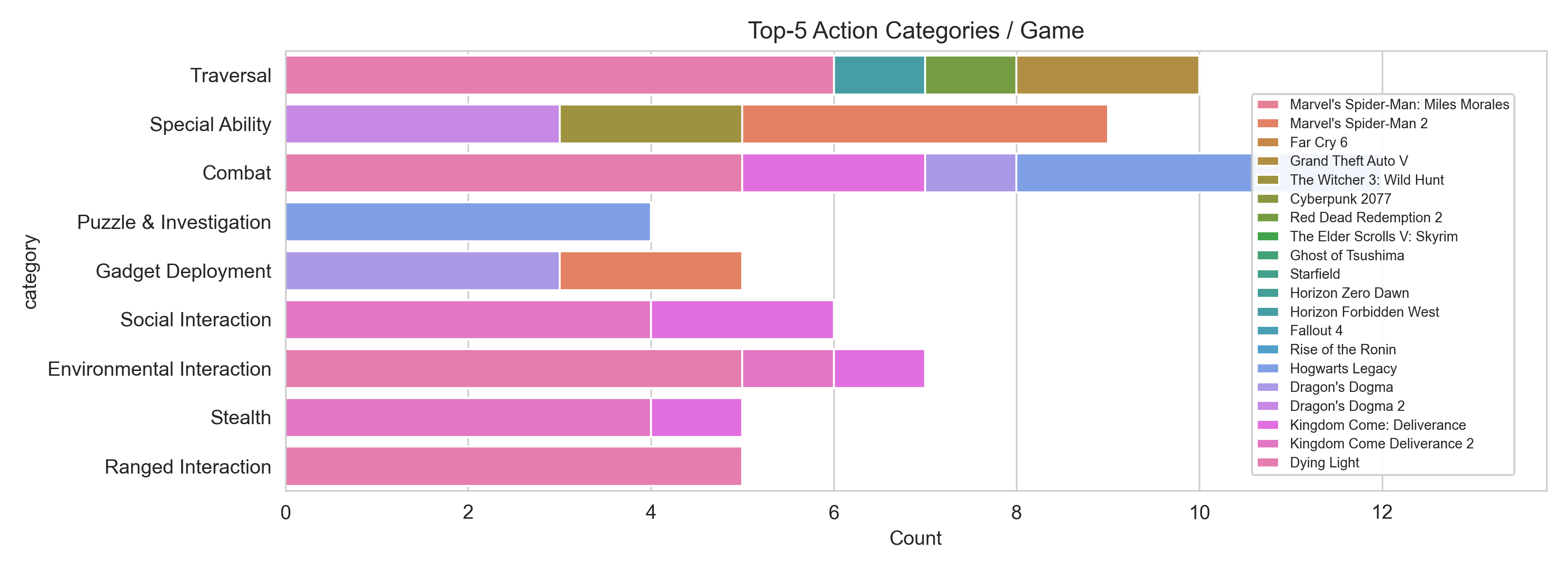}
  \caption{Top action-category counts per game, shown as horizontal stacked bars (longer bars indicate dominant mechanics; shorter bars suggest under-represented ones). To reduce color dependence, the contributing titles for each row (top$\rightarrow$bottom) are: 1) Traversal—\emph{Marvel's Spider-Man: Miles Morales}, \emph{Fallout 4}, \emph{Red Dead Redemption 2}, \emph{Far Cry 6}. 2) Special Ability—\emph{Dragon's Dogma 2}, \emph{The Witcher 3: Wild Hunt}, \emph{Marvel's Spider-Man 2}. 3) Combat—\emph{Marvel's Spider-Man: Miles Morales}, \emph{Kingdom Come: Deliverance} (KCD), \emph{Dragon's Dogma 2}, \emph{Hogwarts Legacy}. 4) Puzzle \& Investigation—\emph{Hogwarts Legacy}. 5) Gadget Deployment—\emph{Dragon's Dogma}, \emph{Marvel's Spider-Man 2}. 6) Social Interaction—\emph{Kingdom Come: Deliverance}, \emph{Kingdom Come: Deliverance 2}. 7) Environmental Interaction—\emph{Kingdom Come: Deliverance}, \emph{Kingdom Come: Deliverance 2}, \emph{Dying Light}. 8) Stealth—\emph{Kingdom Come: Deliverance}, \emph{Kingdom Come: Deliverance 2}. 9) Ranged Interaction—\emph{Dying Light}.}

  \Description{Stacked horizontal bar chart of top action-category counts by game. Y-axis lists action categories; X-axis shows counts. Each bar is stacked by game with a legend naming the contributing titles. Overall, Traversal, Special Ability, and Combat show the largest totals; Environmental and Social Interaction are mid-range; Puzzle \& Investigation, Stealth, Gadget Deployment, and Ranged Interaction are smaller. Longer bars indicate categories that dominate within those games.}

  \label{fig:top-categories}
\end{figure*}

\begin{figure*}[t]
  \centering
  \includegraphics[width=\linewidth]{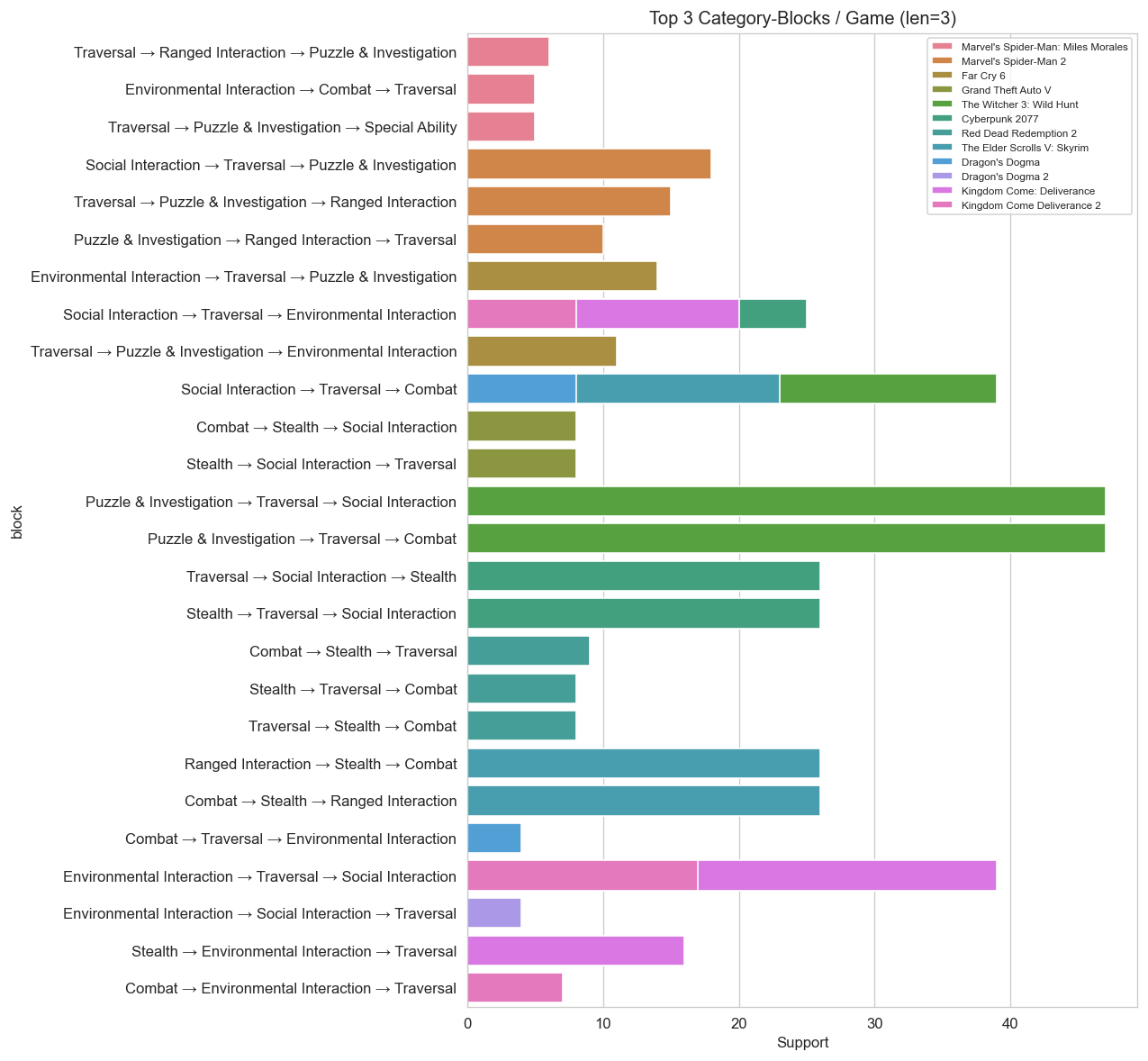}
  \caption{Frequent three-step category blocks (A $\rightarrow$ B $\rightarrow$ C) across games. Horizontal bars show support; rows are the blocks; bars are stacked by game. To reduce color dependence, the contributing games per row (top$\rightarrow$bottom) are, by abbreviation: 1) MM; 2) MM; 3) MM; 4) SM2; 5) SM2; 6) SM2; 7) FC6; 8) KCD, KCD2, CP2077; 9) FC6; 10) DD, Skyrim, TW3; 11) GTA5; 12) GTA5; 13) TW3; 14) TW3; 15) CP2077; 16) CP2077; 17) RDR2; 18) RDR2; 19) RDR2; 20) Skyrim; 21) Skyrim; 22) DD; 23) KCD, KCD2; 24) DD2; 25) KCD; 26) KCD2. Abbrev: MM = \emph{Marvel’s Spider-Man: Miles Morales}, SM2 = \emph{Marvel’s Spider-Man 2}, FC6 = \emph{Far Cry 6}, GTA5 = \emph{Grand Theft Auto V}, TW3 = \emph{The Witcher 3: Wild Hunt}, CP2077 = \emph{Cyberpunk 2077}, RDR2 = \emph{Red Dead Redemption 2}, Skyrim = \emph{The Elder Scrolls V: Skyrim}, DD = \emph{Dragon’s Dogma}, DD2 = \emph{Dragon’s Dogma 2}, KCD = \emph{Kingdom Come: Deliverance}, KCD2 = \emph{Kingdom Come: Deliverance 2}.}
  
  \Description{Stacked horizontal bar chart of frequent three-step action-category blocks by game. X-axis shows support counts; Y-axis lists blocks in the form A to B to C. Each bar is stacked by game with a legend naming the contributing titles. Overall patterns: traversal combined with puzzle/investigation or social interaction appears frequently; several stealth–traversal–combat variants show moderate support; environmental-interaction sequences occur but are less common. The visualization highlights which structural motifs recur and how their support varies across franchises.}

  \label{fig:category-blocks}
\end{figure*}

The Browse mode complements the two in-text exemplars—quality-flow (Figure~\ref{fig:flow}) and action timeline (Figure~\ref{fig:action-timeline}) with several auxiliary panels fed by the same per-mission action sequence and MAQV traces. For transparency, each mission's metadata (title, quest type, and the fixed-snapshot walkthrough text used for extraction) is shown alongside the raw, closed-vocabulary action sequence. This lets designers verify tokenization and mapping decisions at a glance and annotate omissions or misclassifications if needed. A storyboard view renders the same sequence as a linear chain of labeled boxes with arrows; it is produced by collapsing consecutive identical categories and pretty-printing verb names, so repeated ``travel$\rightarrow$combat$\rightarrow$travel'' loops or unusual beats such as ``puzzle before exposition'' surface instantly. A compact numerical summary reports the mean and standard deviation of each MAQV dimension over mission progress; these statistics are computed on raw (unsmoothed) values and serve as a portable fingerprint for spreadsheets or reports. Finally, an action view table (Figure~\ref{fig:action-view}) lists every per-game action with its six MAQV scores and a short description; this table is the reference scaffold behind the extractor and helps relate concrete verbs (e.g., ``Web Zip'') to experiential emphasis (e.g., Uniqueness vs.\ Combat). Across all Browse visuals, colors are shared with the action timeline, and progress is normalized to \([0,100]\%\) so peaks in the quality-flow curves can be mentally aligned to colored timeline segments; when smoothing is used (Gaussian, \(\sigma=2\)), it is strictly for visualization, never for statistics.

The Compare mode aggregates across selected titles and exposes cross-game tendencies using a family of complementary summaries. A combined radar overview places overlaid polygons of mean \emph{mission-level} MAQV scores per title (Figure~\ref{fig:mission-radar}), answering ``which dimensions are emphasized at a glance?'' and allowing immediate contrast of experiential profiles across franchises and sub-genres; values are computed by averaging mission vectors within each game.

Small-multiple radar grids extend this idea at two granularities (the normalized per-game \emph{mission} grid appears in the main text as Figure~\ref{fig:mission-radar-per-game}). One grid aggregates over \emph{actions} (averaging each game's unique action vectors), highlighting how a franchise's verb set is calibrated; the other aggregates over \emph{missions} (averaging per-mission vectors), emphasizing authored flow. A front-end ``Normalize'' toggle rescales each mini-chart to its own maximum so that ``shape'' encodes within-title balance while avoiding visual domination by larger-magnitude titles; switching normalization off restores absolute magnitudes to support scale-aware reading.

Dimensionality-reduction maps translate centroid vectors into spatial metaphors. An Action Similarity Map and a Mission Similarity Map plot principal components computed from per-title action and mission centroids, respectively; clusters and outliers suggest genre affinities and divergences (e.g., why a traversal-heavy modern city title might neighbor another with similar emphasis). Complementary distance matrices visualize the same relationships as colored grids, where each cell encodes pairwise Euclidean distance between titles; darker or lighter tones immediately flag tight siblings and outliers without requiring axis interpretation.

Recurring structure is surfaced with frequency plots derived from sliding windows over the extracted sequences (Figure~\ref{fig:category-blocks}). Frequent category-blocks count support for the top three \(A\!\rightarrow\!B\!\rightarrow\!C\) category patterns per game, exposing shared or franchise-specific loops (e.g., traversal \(\rightarrow\) investigation \(\rightarrow\) ability) computed by scanning the category stream of each mission and tallying three-step windows across the corpus.

A companion frequent action-blocks view repeats the same computation on concrete verbs instead of categories, making signature mechanics visible (e.g., \emph{Web Zip} \(\rightarrow\) Traversal \(\rightarrow\) Combat in superhero titles). Two top-\(k\) bar summaries (one for actions, one for categories) report, per game, the absolute counts of the five most common atoms; long bars imply dominant mechanics while short bars suggest potential design gaps or niches. The \emph{Top-5 Categories / Game} chart is the exemplar retained in the main text (Figure~\ref{fig:top-categories}); the \emph{Top-5 Actions / Game} counterpart follows the same computation and reading logic.

A hierarchical Similarity Tree (Ward dendrogram) clusters titles by their action centroids; branch heights encode distance, yielding an at-a-glance taxonomy of design kinship that complements the PCA scatter without requiring axis literacy. Two centroid tables provide numeric counterparts to the radars: the Action-Centroid Table averages over unique actions per game, and the Mission-Centroid Table averages over missions; each row includes raw means in \([0,1]\) and within-row percentages normalized to that title's highest dimension to enable scale-independent comparison. All Compare summaries are computed on demand from the same cached JSON dataset that feeds Browse, ensuring consistent values when toggling subsets of selected games or exporting snapshots.
\section{Participant Demographics}\label{app:demographics}

\begin{figure*}[t]
  \centering
  \includegraphics[width=\textwidth]{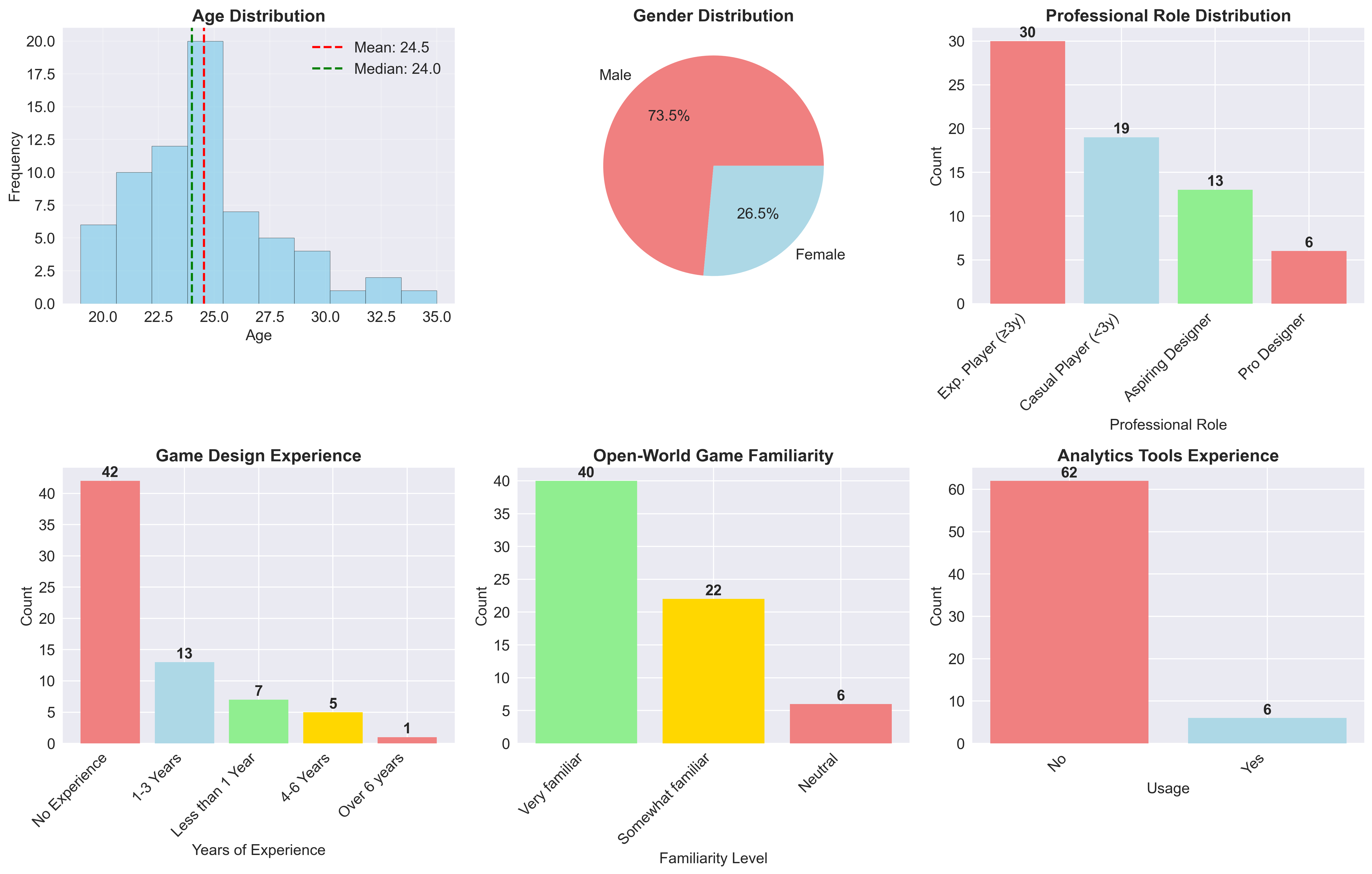}
  \caption{User demographics overview (\(n{=}68\)): age histogram with markers (Mean \(=24.5\); Median \(=25.0\)); gender distribution (73.5\% male, 26.5\% female); professional roles (30 experienced players, 19 casual players, 13 aspiring/indie designers, 6 professional designers); game-design experience (42 none; among those with experience: 13 with 1–3 years, 7 with \(<\)1 year, 5 with 4–6 years, 1 with \(>\)6 years); open-world familiarity (40 very familiar, 22 somewhat familiar, 6 neutral); prior analytics-tool use (62 no, 6 yes).}
  \Description{Participant demographics overview across six panels (n=68). Layout: a 2x3 grid. Top-left: age histogram (x: age in years; y: frequency) with dashed lines indicating mean and median. Top-middle: gender pie chart. Top-right: professional-role bar chart (experienced player, casual player, aspiring/indie designer, professional designer). Bottom-left: game-design experience bar chart (no experience; one to three years; less than one year; four to six years; over six years). Bottom-middle: open-world game familiarity bar chart (very familiar; somewhat familiar; neutral). Bottom-right: prior analytics-tool use bar chart (yes/no). Overall trends: the sample is predominantly male, most participants report no much game-design experience, familiarity with open-world games is high, and prior analytics-tool use is uncommon.}

  \label{fig:user_demographics}
\end{figure*}

In Figure~\ref{fig:user_demographics}, we report the full demographic breakdown of the study sample (\(n{=}68\)). 
Participants were aged 19–35 years (mean \(=24.5\), SD \(=3.2\); median \(=25.0\)). 
Gender: 50 male (73.5\%), 18 female (26.5\%). 
Professional role: 30 experienced players (\(\ge\)3 years; 44.1\%), 19 casual players (\(<\)3 years; 27.9\%), 13 aspiring/indie designers (19.1\%), and 6 professional quest/mission designers (8.8\%). 
Open-world familiarity: 40 very familiar (58.8\%), 22 somewhat familiar (32.4\%), 6 neutral (8.8\%). 
Prior analytics-tool use was rare: 6 yes (8.8\%), 62 no (91.2\%). 
Regarding game-design experience, 42 reported no experience (61.8\%). 
Among the 26 with any design experience, 13 had 1–3 years (50.0\%), 7 had \(<\)1 year (26.9\%), and 5 had 4–6 years (19.2\%); 1 reported \(>\)6 years (3.8\%).

\section{Stage 2 Reflection Prompts (Full)}\label{app:stage2-prompts}

Participants were shown the following 12 prompts and instructed to select any $\geq 3$ to answer. The abridged versions appear in Table~\ref{tab:stage2}.

\begin{enumerate}[label=\textbf{Q\arabic*}, leftmargin=*, itemsep=0.6em]

\item \label{q:radar-pointcloud}
Using the six-dimension radar and point-cloud plots, summarize the commonalities and divergences across games. Which dimensions define each game’s signature profile?

\item \label{q:good-vs-flawed-centers}
Select one well-designed and one flawed mission. From their dimensional distributions alone, what centers of gravity or voids distinguish them?

\item \label{q:action-diversity-block-structure}
Based on the action point cloud and frequent action-block charts, describe the patterns of action diversity and block structure across games.

\item \label{q:good-poor-shared-missing-combos}
For your chosen good vs.\ poor missions, compare only the action-block sequences: which block combinations are shared or missing?

\item \label{q:long-missions-pacing}
Analyze several long missions. How do well-constructed vs.\ poorly-constructed long missions differ in dimensional centers, block sequencing, and pacing?

\item \label{q:neglect-overemphasis}
Across the games you explored, which dimensions or categories are neglected or over-emphasized? How might this imbalance create pacing gaps or content hollowness?

\item \label{q:main-side-poi}
Within one game you know well, contrast Main, Side, and POI missions in dimensional layout and action flow. What design priorities emerge?

\item \label{q:series-evolution}
Examine multiple instalments of the same IP or similar titles. Describe mission/action evolution and whether it enriched or diminished your experience.

\item \label{q:formula-vs-monotony}
Which missions/games break the formula and exhibit uniqueness, and which fall into monotony due to rigid templates? Explain how you judged.

\item \label{q:ingredients}
Outline the core ingredients and flavouring ingredients of a successful open-world mission formula based on your observations.

\item \label{q:motives-mapping}
Map the six dimensions onto player motives (challenge, story, exploration, immersion). Which combos satisfy multiple motives? Give examples.

\item \label{q:peaks-valleys}
In the peak-valley curves, which dimensions or blocks serve as emotional climaxes or recovery buffers? What pacing strategies does this suggest?

\end{enumerate}

\section{Extended Distributions}\label{app:extended-plots}

To complement the main-text summaries (e.g., Table~\ref{tab:quantSummary}), this appendix provides the UEQ-S overall plots, and per-item distribution plots for Stage 1 validity items (A1, A2, A5–A7), the System Usability Scale (SUS1–SUS10), and the UEQ-S (U1–U8), with overlaid means and medians to substantiate the aggregate statistics.

\begin{figure*}[t]
  \centering
  \includegraphics[width=0.95\linewidth]{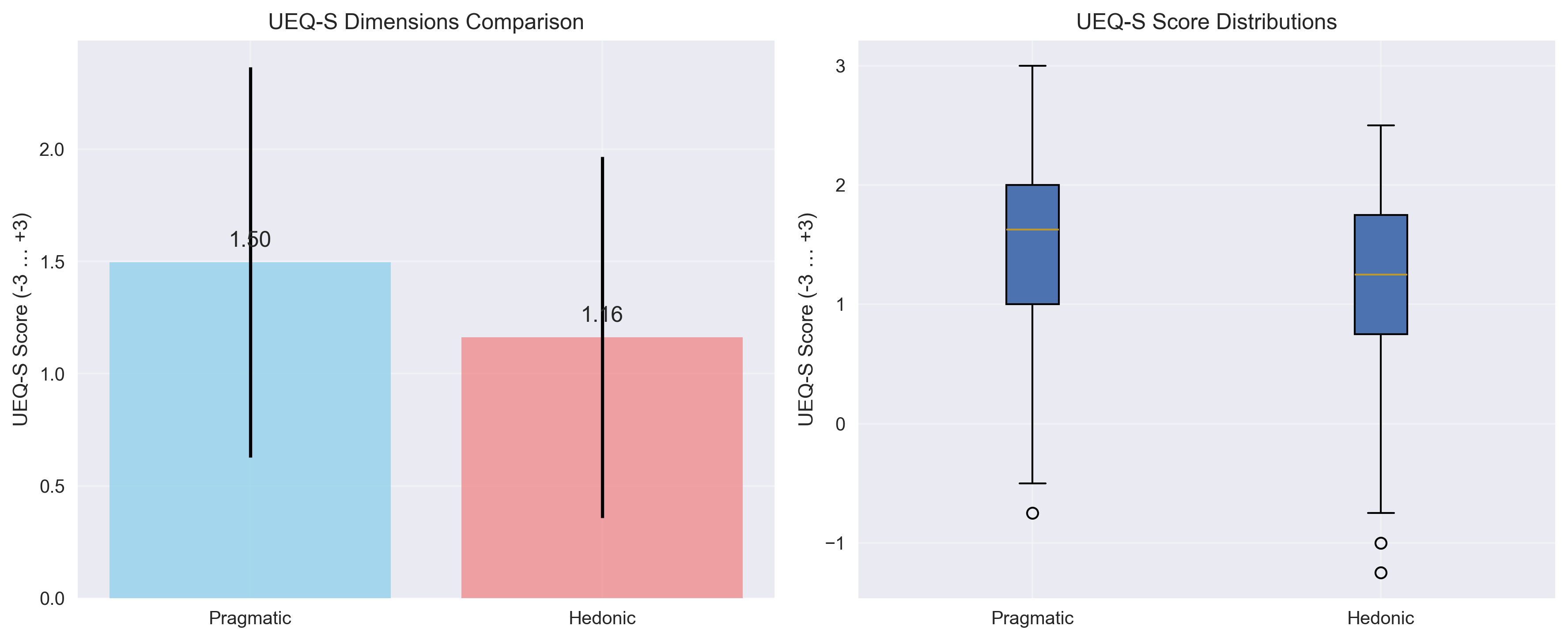}
  \caption{UEQ-S analysis. Both pragmatic and hedonic dimensions are positive: pragmatic mean $=1.50$ (median $=1.62$), hedonic mean $=1.16$ (median $=1.25$), on the $-3\ldots+3$ scale.}
  \Description
  {UEQ-S pragmatic and hedonic scores positive, centered above 1. Two panels for $n=68$: (left) bar chart with error whiskers shows pragmatic dimension mean 1.50 and hedonic mean 1.16, both above zero; (right) boxplots show both distributions centered above 1, with pragmatic slightly higher and tighter, hedonic a bit lower with a few mild negative outliers. Overall, both dimensions are clearly on the positive side, indicating favorable perceptions.}
  \label{fig:ueq-analysis}
\end{figure*}

\begin{figure*}[t]
  \centering
  \includegraphics[width=\textwidth]{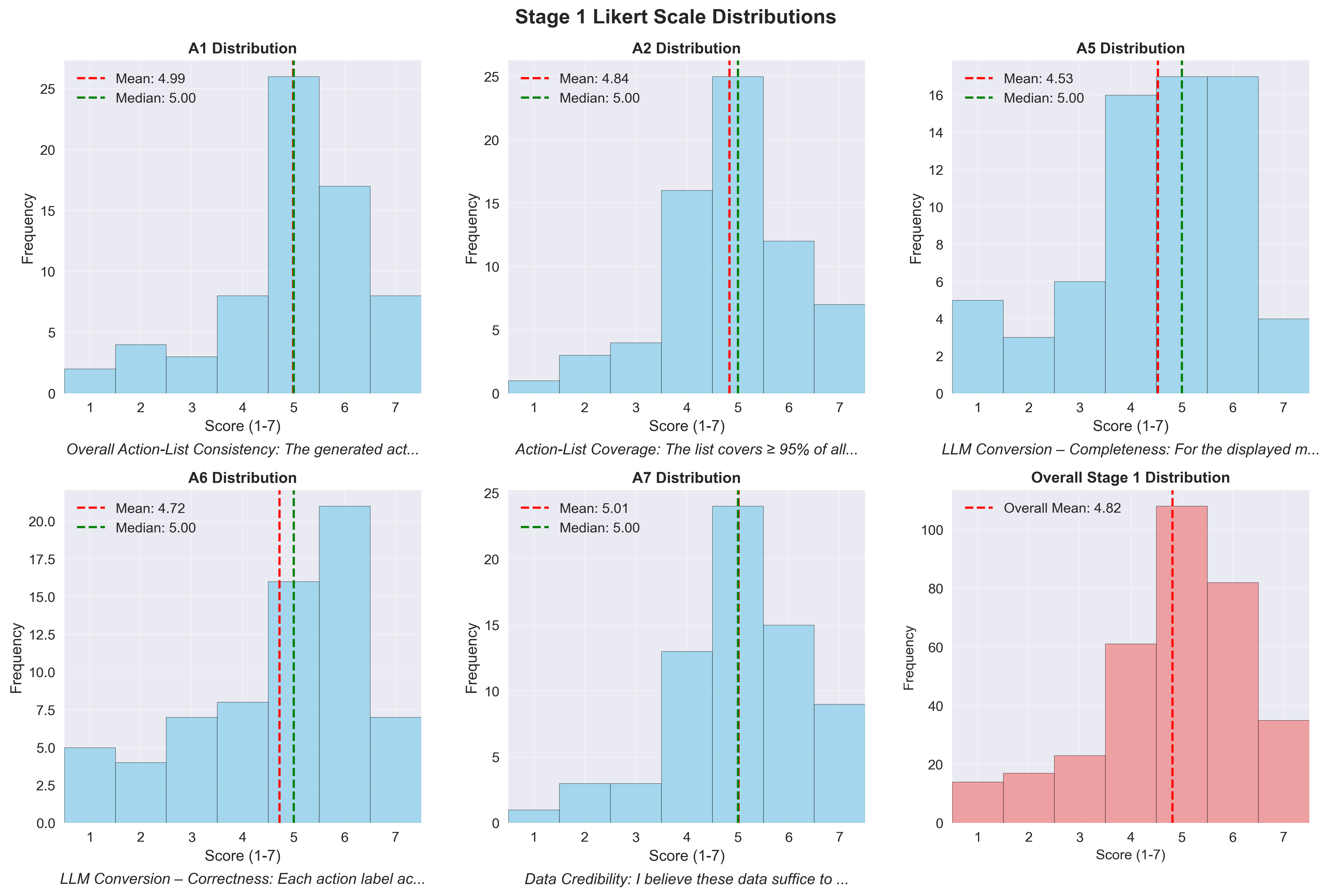}
  \caption{Stage~1 per-item distributions (Likert 1–7) with dashed lines marking the mean (red) and median (green). Distributions cluster around 5; the overall mean across A1, A2, A5–A7 is \(4.82\). Per-item means: A1 \(=4.99\), A2 \(=4.84\), A5 \(=4.53\), A6 \(=4.72\), A7 \(=5.01\) (all medians \(=5.00\)).}
  \Description{Six-panel histograms of Stage~1 Likert scores with mean and median markers. Layout: a 2×3 grid. Each panel shows a histogram on a 1–7 Likert scale (x: score; y: frequency) with vertical dashed lines indicating the mean and median. Panels correspond to A1 (Action-List Consistency), A2 (Action-List Coverage), A5 (LLM Completeness), A6 (LLM Correctness), A7 (Data Credibility), and the overall aggregate. Overall trends: all per-item distributions are unimodal and cluster near 5; tails at very low or very high scores are light; the aggregate histogram centers near 5 without ceiling or floor effects. Exact means and medians appear in the caption.}

  \label{fig:stage1-dists}
\end{figure*}

\begin{figure*}[t]
  \centering
  \includegraphics[width=\textwidth]{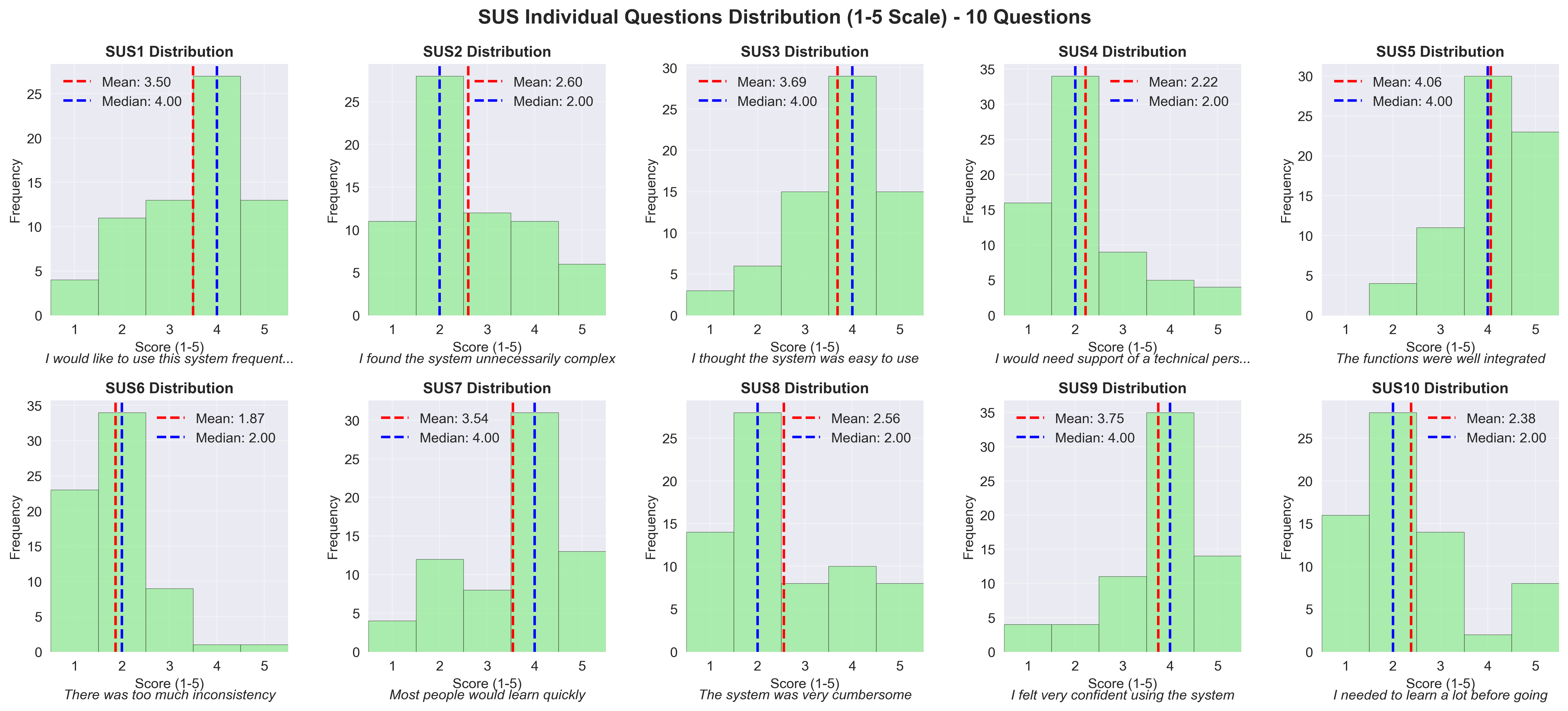}
  \caption{SUS per-item distributions (1–5 scale). Dashed lines mark the mean (red) and median (blue). Positively worded items (SUS1, 3, 5, 7, 9) skew higher, while negatively worded items (SUS2, 4, 6, 8, 10) skew lower; this polarity pattern is consistent with the aggregate SUS computed from these items (see Table~\ref{tab:quantSummary}).}
  \Description{Ten-panel histograms of SUS item scores with mean and median markers (n=68). Layout: a 2×5 grid. Each panel shows a 1–5 Likert histogram (x: score; y: frequency) with dashed lines indicating the mean and median. Polarity pattern: positively worded items (SUS1, 3, 5, 7, 9) concentrate at higher scores, while negatively worded items (SUS2, 4, 6, 8, 10) concentrate at lower scores. Distributions are unimodal with no ceiling or floor effects.}

  \label{fig:sus-indiv}
\end{figure*}

\begin{figure*}[t]
  \centering
  \includegraphics[width=\textwidth]{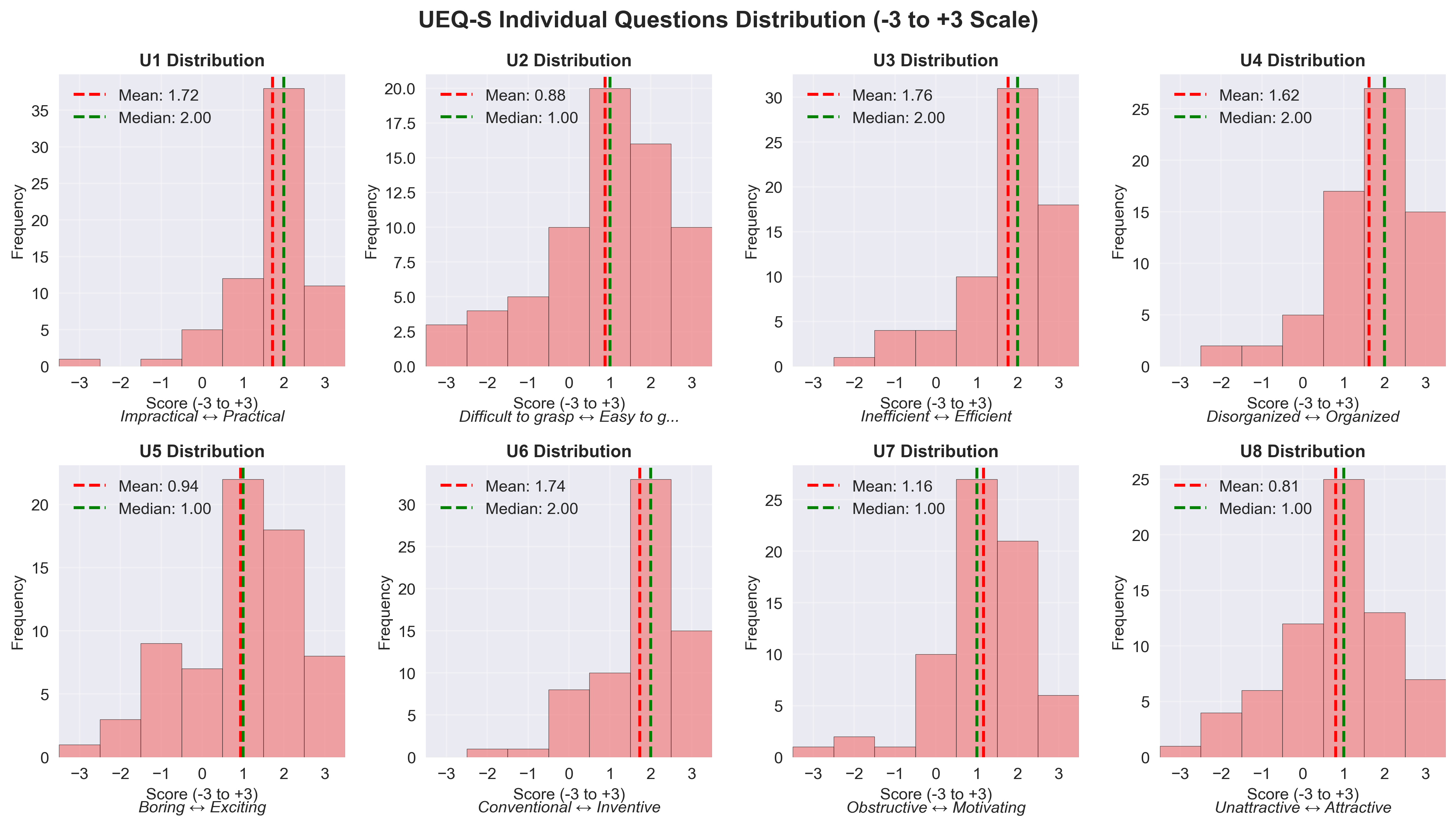}
  \caption{UEQ-S per-item distributions on the \(-3\) to \(+3\) scale. Dashed lines mark the mean (red) and median (green). All items are shifted to the positive side; the aggregated UEQ-S means are \(1.50\) (Pragmatic: U1–U4) and \(1.16\) (Hedonic: U5–U8).}
  \Description{Eight-panel histograms of UEQ-S item scores with mean and median markers (n=68). Layout: a 2×4 grid. Each panel shows a histogram on the −3 to +3 semantic-differential scale (x: score; y: frequency) with vertical dashed lines indicating the mean and median. Items U1–U4 represent pragmatic quality (Impractical to Practical; Difficult to grasp to Easy to grasp; Inefficient to Efficient; Disorganized to Organized). Items U5–U8 represent hedonic quality (Boring to Exciting; Conventional to Inventive; Obstructive to Motivating; Unattractive to Attractive). Overall trends: all items are shifted to the positive side; pragmatic items tend to be slightly higher than hedonic items; distributions are unimodal without pronounced negative tails. Exact aggregated means appear in the caption.}

  \label{fig:ueq-indiv}
\end{figure*}

\section{Pre-annotation IRR for MAQV}
\label{app:irr}

Before the full annotation pass, two raters (two authors) independently scored actions from two representative titles (The Witcher 3, GTA V) to check inter-rater reliability (IRR). For this IRR only, each MAQV dimension was discretized to $\{0, 0.25, 0.5, 0.75, 1\}$ to improve comparability. We report quadratic-weighted Cohen's $\kappa$ with 2{,}000 bootstrap resamples (seed$=$42), together with Spearman's $\rho$, exact/off-by-one rates, and mean absolute difference (mapped back to $[0,1]$). Main-paper analyses revert to continuous 0--1 scores and the structured consensus in Section~\ref{sec:actionblock}.

\begin{table}[t]
\centering
\small
\setlength{\tabcolsep}{4pt}
\caption{Pre-annotation IRR on discretized MAQV scores (two games, 76 actions).}
\label{tab:irr_per_dim_full}
\begin{tabular}{@{}lcccccc@{}}
\toprule
Dim & $\kappa_w$ & CI & $\rho$ & Exact & Off$\pm$1 & MAD \\
\midrule
U & 0.7366 & [0.611, 0.813] & 0.7966 & 0.6184 & 0.3816 & 0.0954 \\
C & 0.9658 & [0.946, 0.982] & 0.9682 & 0.8553 & 0.1447 & 0.0362 \\
N & 0.8682 & [0.746, 0.931] & 0.9077 & 0.7632 & 0.2237 & 0.0625 \\
E & 0.9530 & [0.918, 0.977] & 0.9536 & 0.8553 & 0.1447 & 0.0362 \\
P & 0.8904 & [0.834, 0.934] & 0.9218 & 0.7895 & 0.2105 & 0.0526 \\
A & 0.8434 & [0.747, 0.908] & 0.8439 & 0.7237 & 0.2632 & 0.0724 \\
\midrule
Overall & 0.9131 & [0.895, 0.929] & 0.9134 & 0.7675 & 0.2281 & 0.0592 \\
\bottomrule
\end{tabular}
\end{table}

Three patterns in Table~\ref{tab:irr_all_items} support the reasonableness of the ratings. First, scores align with construct expectations for both raters—core combat actions are high on C, traversal/vehicle actions on E, dialogue/calls on N, and signature abilities on U/A. Second, disagreement is small: 74/76 actions have a maximum step difference $\leq 1$ on the 0–4 ordinal grid, with only two reaching 2 (GTAV: \emph{Stock Market Trade}, \emph{Parachute Free-fall}), consistent with $\kappa_w{=}0.913$ and exact agreement $0.768$. Third, within-category stability is strong (e.g., multiple ranged or vehicle actions show near-identical profiles across raters), indicating shared decision rules rather than idiosyncratic judgments. Overall, the table demonstrates that the MAQV protocol is interpretable and replicable, with residual differences confined to boundary cases (typically novelty or affect), not construct confusion.

\onecolumn
\setlength{\tabcolsep}{3pt}
\small
\begin{longtable}{@{}lllrrrrrrrrrrrrr@{}}
\caption{All actions with both raters’ scores (two games, 76 actions).
Abbreviations: TW3 = The Witcher 3;
Trav = Traversal, Cbt = Combat, Env = Environmental Interaction, Gdt = Gadget Deployment,
PI = Puzzle \& Investigation, Rng = Ranged Interaction, Soc = Social Interaction,
SpA = Special Ability, Stl = Stealth.
Columns list rater A/B scores on each MAQV dimension (0–1 grid); max$\Delta$ is the
maximum absolute step difference (0–4) across the six dimensions for that action.}

\label{tab:irr_all_items}\\
\toprule
Game & Category & Action & U$_A$ & U$_B$ & C$_A$ & C$_B$ & N$_A$ & N$_B$ & E$_A$ & E$_B$ & P$_A$ & P$_B$ & A$_A$ & A$_B$ & max$\Delta$ \\
\midrule
\endfirsthead
\toprule
Game & Category & Action & U$_A$ & U$_B$ & C$_A$ & C$_B$ & N$_A$ & N$_B$ & E$_A$ & E$_B$ & P$_A$ & P$_B$ & A$_A$ & A$_B$ & max$\Delta$ \\
\midrule
\endhead
\midrule
\multicolumn{16}{r}{(continued)}\\
\endfoot
\bottomrule
\endlastfoot
GTA V & Cbt & Assault-Rifle Burst & 0.25 & 0.25 & 1 & 1 & 0 & 0 & 0 & 0 & 0 & 0 & 0.75 & 0.75 & 0 \\
GTA V & Cbt & Franklin Driving Focus & 0.75 & 1 & 0.5 & 0.75 & 0 & 0 & 0.5 & 0.5 & 0.25 & 0 & 0.75 & 1 & 1 \\
GTA V & Cbt & Grenade Throw & 0.25 & 0.25 & 0.75 & 0.75 & 0 & 0 & 0 & 0 & 0.25 & 0.25 & 0.75 & 0.75 & 0 \\
GTA V & Cbt & Melee Punch Combo & 0 & 0.25 & 0.5 & 0.5 & 0 & 0 & 0 & 0 & 0 & 0 & 0.5 & 0.25 & 1 \\
GTA V & Cbt & Michael Bullet Time & 0.75 & 1 & 0.75 & 0.75 & 0 & 0 & 0 & 0 & 0.25 & 0 & 0.75 & 1 & 1 \\
GTA V & Cbt & Pistol Fire & 0.25 & 0.25 & 1 & 0.75 & 0 & 0 & 0 & 0 & 0 & 0 & 0.75 & 0.5 & 1 \\
GTA V & Cbt & Rocket Launcher & 0.25 & 0.5 & 1 & 1 & 0 & 0 & 0 & 0 & 0 & 0 & 1 & 1 & 0 \\
GTA V & Cbt & SMG Spray & 0.25 & 0.25 & 1 & 0.75 & 0 & 0 & 0 & 0 & 0 & 0 & 0.75 & 0.5 & 1 \\
GTA V & Cbt & Sniper Shot & 0.25 & 0.25 & 1 & 0.75 & 0 & 0 & 0 & 0 & 0.25 & 0.25 & 0.75 & 0.75 & 0 \\
GTA V & Cbt & Trevor Rampage & 0.75 & 1 & 1 & 1 & 0 & 0 & 0 & 0 & 0 & 0 & 1 & 1 & 0 \\
GTA V & Env & ATM Withdrawal & 0 & 0 & 0 & 0 & 0 & 0 & 0 & 0 & 0 & 0 & 0 & 0 & 0 \\
GTA V & Env & Hack Security Keypad & 0.25 & 0.5 & 0 & 0.25 & 0.25 & 0.25 & 0 & 0 & 1 & 0.75 & 0.5 & 0.5 & 1 \\
GTA V & Env & Loot Cash Register & 0.25 & 0.25 & 0.25 & 0.25 & 0.25 & 0 & 0 & 0 & 0 & 0 & 0.5 & 0.5 & 0 \\
GTA V & Env & Open Safe & 0.25 & 0.5 & 0 & 0.25 & 0.25 & 0.25 & 0 & 0 & 0.75 & 0.5 & 0.5 & 0.5 & 1 \\
GTA V & Gdt & Drone Recon (SP Heist) & 0.5 & 0.5 & 0.25 & 0.25 & 0.5 & 0.25 & 0.5 & 0.5 & 0.5 & 0.5 & 0.5 & 0.5 & 1 \\
GTA V & Gdt & Remote-Sentry Turret (Heist) & 0.5 & 0.5 & 0.75 & 0.75 & 0.25 & 0.25 & 0 & 0 & 0.5 & 0.5 & 0.75 & 0.75 & 0 \\
GTA V & PI & Dossier Search (PC) & 0.25 & 0.25 & 0 & 0 & 0.75 & 0.5 & 0 & 0 & 0.75 & 0.75 & 0.25 & 0.25 & 1 \\
GTA V & PI & Photo Evidence (Cell) & 0.25 & 0.25 & 0 & 0 & 0.75 & 0.5 & 0.25 & 0 & 0.5 & 0.5 & 0.25 & 0.25 & 1 \\
GTA V & Rng & Flare Gun & 0.25 & 0.5 & 0.25 & 0.25 & 0.25 & 0.25 & 0 & 0 & 0.25 & 0.25 & 0.25 & 0.5 & 1 \\
GTA V & Rng & Sticky Bomb Placement & 0.25 & 0.5 & 0.75 & 0.75 & 0 & 0 & 0 & 0 & 0.5 & 0.5 & 0.75 & 0.75 & 1 \\
GTA V & Rng & Tear Gas Throw & 0.25 & 0.25 & 0.5 & 0.5 & 0 & 0 & 0 & 0 & 0.25 & 0.25 & 0.5 & 0.5 & 0 \\
GTA V & Soc & Cell-Phone Call & 0.25 & 0.25 & 0 & 0 & 1 & 0.75 & 0 & 0 & 0.25 & 0.25 & 0.25 & 0.25 & 1 \\
GTA V & Soc & Dialogue Choice (Heist Crew) & 0.5 & 0.5 & 0 & 0 & 1 & 1 & 0 & 0 & 0.5 & 0.5 & 0.5 & 0.5 & 0 \\
GTA V & Soc & Stock Market Trade & 0.25 & 0.5 & 0 & 0 & 0.5 & 0 & 0 & 0 & 0.5 & 0.5 & 0.25 & 0.25 & 2 \\
GTA V & SpA & Director Mode & 0.75 & 0.75 & 0 & 0 & 0 & 0.25 & 0 & 0.25 & 0.25 & 0.25 & 0.25 & 0.25 & 1 \\
GTA V & Stl & Silenced Pistol Shot & 0.25 & 0.25 & 0.75 & 0.75 & 0.25 & 0 & 0 & 0 & 0.25 & 0 & 0.5 & 0.5 & 1 \\
GTA V & S & Stealth Takedown & 0.25 & 0.25 & 0.75 & 0.75 & 0.25 & 0 & 0 & 0 & 0.25 & 0.25 & 0.75 & 0.5 & 1 \\
GTA V & S & Stealth Walk & 0.25 & 0.25 & 0.25 & 0.25 & 0.25 & 0 & 0.25 & 0.25 & 0.25 & 0.25 & 0.5 & 0.5 & 0 \\
GTA V & Trav & Bicycle Pedal & 0.25 & 0.25 & 0 & 0 & 0 & 0 & 0.75 & 0.5 & 0 & 0 & 0.25 & 0.25 & 1 \\
GTA V & Trav & Boat Driving & 0.25 & 0.25 & 0 & 0 & 0 & 0 & 0.75 & 0.75 & 0 & 0 & 0.25 & 0.25 & 0 \\
GTA V & Trav & Car Driving & 0.25 & 0.25 & 0 & 0 & 0 & 0 & 1 & 1 & 0 & 0 & 0.25 & 0.25 & 0 \\
GTA V & Trav & Fixed-Wing Pilot & 0.25 & 0.5 & 0 & 0 & 0 & 0 & 1 & 1 & 0 & 0 & 0.5 & 0.75 & 1 \\
GTA V & Trav & Helicopter Pilot & 0.25 & 0.5 & 0 & 0 & 0 & 0 & 1 & 1 & 0 & 0 & 0.5 & 0.75 & 1 \\
GTA V & Trav & Motorcycle Riding & 0.25 & 0.25 & 0 & 0 & 0 & 0 & 1 & 0.75 & 0 & 0 & 0.25 & 0.5 & 1 \\
GTA V & Trav & Parachute Free-fall & 0.5 & 0.5 & 0 & 0 & 0 & 0 & 0.5 & 0.75 & 0 & 0 & 0.5 & 1 & 2 \\
GTA V & Trav & Scuba Dive & 0.25 & 0.5 & 0 & 0 & 0 & 0 & 0.5 & 0.75 & 0.25 & 0 & 0.25 & 0.5 & 1 \\
GTA V & Trav & Sprint & 0 & 0.25 & 0 & 0 & 0 & 0 & 0.5 & 0.5 & 0 & 0 & 0.25 & 0.25 & 1 \\
GTA V & Trav & Walk / Jog & 0 & 0.25 & 0 & 0 & 0 & 0 & 0.5 & 0.5 & 0 & 0 & 0 & 0 & 1 \\
TW3 & Cbt & Bomb Throw (Grapeshot) & 0.25 & 0.5 & 0.75 & 0.75 & 0 & 0 & 0 & 0 & 0.25 & 0.25 & 0.75 & 0.75 & 1 \\
TW3 & Cbt & Fast Dodge / Roll & 0.25 & 0.25 & 0.5 & 0.5 & 0 & 0 & 0 & 0.25 & 0 & 0 & 0.5 & 0.5 & 0 \\
TW3 & Cbt & Parry / Riposte & 0.25 & 0.25 & 0.75 & 0.75 & 0 & 0 & 0 & 0 & 0 & 0 & 0.5 & 0.5 & 0 \\
TW3 & Cbt & Silver Sword Combo & 0.5 & 0.5 & 1 & 1 & 0 & 0 & 0 & 0 & 0 & 0 & 0.75 & 0.5 & 1 \\
TW3 & Cbt & Steel Sword Combo & 0.25 & 0.25 & 1 & 1 & 0 & 0 & 0 & 0 & 0 & 0 & 0.75 & 0.5 & 1 \\
TW3 & Cbt & Strong Attack & 0.25 & 0.25 & 1 & 0.75 & 0 & 0 & 0 & 0 & 0 & 0 & 0.75 & 0.5 & 1 \\
TW3 & Env & Herb Gathering & 0.25 & 0.25 & 0 & 0 & 0 & 0 & 0.5 & 0.25 & 0.25 & 0.25 & 0 & 0 & 1 \\
TW3 & Env & Monster Nest Destruction & 0.25 & 0.5 & 0.25 & 0.25 & 0 & 0.25 & 0 & 0.25 & 0.5 & 1 & 0.25 & 0.5 & 1 \\
TW3 & Env & Torch Ignite / Extinguish & 0.25 & 0.25 & 0 & 0 & 0 & 0 & 0.25 & 0.25 & 0.25 & 0.25 & 0 & 0 & 0 \\
TW3 & Gdt & Dancing Star Bomb & 0.25 & 0.5 & 0.75 & 0.75 & 0 & 0 & 0 & 0 & 0.25 & 0.25 & 0.75 & 0.75 & 1 \\
TW3 & Gdt & Hanged Man's Venom Oil & 0.5 & 0.5 & 0.5 & 0.5 & 0 & 0 & 0 & 0 & 0.5 & 0.5 & 0.5 & 0.25 & 1 \\
TW3 & Gdt & Yrden Trap Glyph & 0.5 & 0.5 & 0.5 & 0.5 & 0 & 0 & 0 & 0 & 0.5 & 0.25 & 0.5 & 0.5 & 1 \\
TW3 & PI & Code Riddle Solve & 0.25 & 0.5 & 0 & 0 & 0.5 & 0.25 & 0 & 0 & 1 & 0.75 & 0.25 & 0.25 & 1 \\
TW3 & PI & Inspect Clue & 0.25 & 0.25 & 0 & 0 & 0.75 & 0.5 & 0 & 0 & 0.75 & 0.5 & 0.25 & 0.25 & 1 \\
TW3 & PI & Track Footprints & 0.25 & 0.5 & 0 & 0 & 0.75 & 0.5 & 0.5 & 0.25 & 0.75 & 0.5 & 0.25 & 0.25 & 1 \\
TW3 & PI & Witcher Sense Track & 0.5 & 0.75 & 0 & 0 & 0.75 & 0.5 & 0.25 & 0.25 & 0.75 & 0.5 & 0.25 & 0.25 & 1 \\
TW3 & Rng & Crossbow Bolt & 0.25 & 0.25 & 0.5 & 0.5 & 0 & 0 & 0 & 0 & 0 & 0 & 0.25 & 0.25 & 0 \\
TW3 & Rng & Northern Wind Bomb & 0.25 & 0.5 & 0.5 & 0.5 & 0 & 0 & 0 & 0 & 0 & 0 & 0.5 & 0.5 & 1 \\
TW3 & Soc & Contract Negotiation & 0.5 & 0.5 & 0 & 0 & 0.75 & 0.5 & 0 & 0 & 0.5 & 0.25 & 0.5 & 0.25 & 1 \\
TW3 & Soc & Dialogue Choice & 0.25 & 0.25 & 0 & 0 & 1 & 1 & 0 & 0 & 0.25 & 0.25 & 0.5 & 0.5 & 0 \\
TW3 & Soc & Gwent Card Game & 0.5 & 0.75 & 0 & 0 & 0.25 & 0.25 & 0 & 0 & 0.75 & 0.75 & 0.25 & 0.5 & 1 \\
TW3 & Soc & Trade with Merchant & 0.25 & 0.25 & 0 & 0 & 0.25 & 0.25 & 0 & 0 & 0.25 & 0.25 & 0 & 0 & 0 \\
TW3 & SpA & Adrenaline Rush (Mutation) & 0.75 & 0.5 & 0.75 & 0.75 & 0 & 0 & 0 & 0 & 0 & 0 & 0.75 & 0.75 & 1 \\
TW3 & SpA & Sign: Aard & 0.75 & 0.75 & 0.75 & 0.5 & 0 & 0 & 0 & 0 & 0.25 & 0.25 & 0.75 & 0.5 & 1 \\
TW3 & SpA & Sign: Axii (Combat Stun) & 0.75 & 0.75 & 0.75 & 0.5 & 0 & 0 & 0 & 0 & 0.25 & 0.25 & 0.5 & 0.5 & 1 \\
TW3 & SpA & Sign: Igni & 0.75 & 0.75 & 0.75 & 0.75 & 0 & 0 & 0 & 0 & 0.25 & 0.25 & 0.75 & 0.75 & 0 \\
TW3 & SpA & Sign: Quen & 0.75 & 0.75 & 0.5 & 0.25 & 0 & 0 & 0 & 0 & 0.25 & 0.25 & 0.5 & 0.5 & 1 \\
TW3 & SpA & Sign: Yrden & 0.75 & 0.75 & 0.5 & 0.5 & 0 & 0 & 0 & 0 & 0.5 & 0.25 & 0.5 & 0.5 & 1 \\
TW3 & S & Crossbow Silent Shot & 0.25 & 0.25 & 0.5 & 0.75 & 0.25 & 0 & 0 & 0 & 0.25 & 0 & 0.5 & 0.5 & 1 \\
TW3 & S & Quiet Approach & 0.25 & 0.25 & 0.25 & 0.25 & 0.25 & 0 & 0.25 & 0.25 & 0.25 & 0.25 & 0.5 & 0.5 & 0 \\
TW3 & Trav & Boat Sailing & 0.25 & 0.5 & 0 & 0 & 0 & 0 & 0.75 & 0.75 & 0 & 0 & 0.25 & 0.25 & 1 \\
TW3 & Trav & Climb / Vault & 0.25 & 0.25 & 0 & 0 & 0 & 0 & 0.5 & 0.25 & 0.25 & 0 & 0.25 & 0.25 & 1 \\
TW3 & Trav & Dive \& Loot & 0.25 & 0.5 & 0 & 0 & 0 & 0 & 0.5 & 0.5 & 0.25 & 0 & 0.25 & 0.5 & 1 \\
TW3 & Trav & Fast-Travel Signpost & 0.5 & 0.25 & 0 & 0 & 0 & 0 & 1 & 1 & 0 & 0 & 0 & 0 & 1 \\
TW3 & Trav & Roach Gallop & 0.5 & 0.5 & 0 & 0 & 0 & 0 & 0.75 & 0.75 & 0 & 0 & 0.25 & 0.25 & 0 \\
TW3 & Trav & Sprint / Stamina Run & 0 & 0.25 & 0 & 0 & 0 & 0 & 0.5 & 0.5 & 0 & 0 & 0.25 & 0.25 & 1 \\
TW3 & Trav & Swim Surface & 0.25 & 0.25 & 0 & 0 & 0 & 0 & 0.5 & 0.5 & 0 & 0 & 0.25 & 0.25 & 0 \\
TW3 & Trav & Walk / Jog & 0 & 0.25 & 0 & 0 & 0 & 0 & 0.5 & 0.5 & 0 & 0 & 0 & 0 & 1 \\
\end{longtable}
\twocolumn

\end{document}